\documentclass[useAMS,usenatbib]{mn2e}
\usepackage{graphicx}
\usepackage{subfigure}
\usepackage{ amssymb }
\usepackage{fmtcount}
\newcommand{\gtsim}{\mbox{{\raisebox{-0.4ex}{$\stackrel{>}{{\scriptstyle\sim}}
$}}}}

\def\Msun{\hbox{$\rm\thinspace M_{\odot}$}}
\def\asec{$^{\prime\prime}$}
\title[Spectrocopic confirmation of luminous LBGs at $z\geq6$]{A remarkably high fraction of strong Ly$\bmath{\alpha}$ emitters amongst luminous redshift $\bmath{6.0<z<6.5}$ Lyman break galaxies in the UKIDSS Ultra-Deep Survey}
\author[E. Curtis-Lake et al.]
{E. Curtis-Lake$^{1}$\thanks{Email: efcl@roe.ac.uk},
R. J. McLure$^{1}$, H. J. Pearce$^{1}$, J. S. Dunlop$^{1}$, M. Cirasuolo$^{1}$, \and
D. P. Stark$^{2}$\thanks{Hubble Fellow}, O. Almaini$^{3}$, E. J. Bradshaw$^{3}$, R. Chuter$^{3}$, S. Foucaud$^{4}$, 
\and W. G. Hartley$^{3}$\\
$^{1}$ SUPA\thanks{Scottish Universities Physics Alliance},
Institute for Astronomy, University of Edinburgh, 
Royal Observatory, Edinburgh EH9 3HJ\\
$^{2}$ Department of Astronomy, Steward Observatory, University of Arizona, 933 North Cherry Avenue, Rm N204, Tucson, AZ, 8572\\
$^{3}$ School of Physics \& Astronomy, University of Nottingham, University Park, Nottingham NG7 2RD\\
$^{4}$ Department of Earth Sciences, National Taiwan Normal University, No. 88, Section 4, Tingzhou Road,\\
{  } Wenshan District, Taipei 11677, Taiwan}

\begin{document}


\pagerange{\pageref{firstpage}--\pageref{lastpage}} \pubyear{2002}

\maketitle

\label{firstpage}

\label{firstpage}

\def\Msun{\hbox{$\rm\thinspace M_{\odot}$}}

\begin{abstract}
We present spectroscopic confirmation of ten highly luminous ($L\geq 2L^{\star}$)
Lyman alpha emitters in the redshift range $6.01<z<6.49$ (nine
galaxies and one AGN), initially drawn from a sample of fourteen $z_{phot}\geq 6$ Lyman 
break galaxies (LBGs) selected from an area of 0.25 square degrees within the UKIDSS Ultra-deep Survey (UDS). 
Overall, our high rate of spectroscopic confirmation ($\geq71$\%) and low rate of contamination 
provides a strong vindication of the photometric redshift analysis used to define the original sample. 
By considering star-formation rate estimates based on the Ly$\alpha$ and UV continuum 
luminosity we conclude that our sample is consistent with a Ly$\alpha$ escape fraction of $\simeq 25\%$.
Moreover, after careful consideration of the potential uncertainties and biases, we find that $40\%-50\%$ of 
our sample of $L\geq 2L^{\star}$ galaxies at $6.0<z<6.5$ display strong Ly$\alpha$ emission (rest-frame equivalent width $\geq25$\AA), 
a fraction which is a factor of $\simeq 2$ higher than previously reported for $L\leq L^{\star}$ galaxies at $z\simeq 6$.
Our results suggest that, as the epoch of reionization is approached, it is plausible that the Ly$\alpha$ emitter fraction amongst luminous ($L\geq 2 L^{\star}$) LBGs shows a similarly sharp increase to that observed in their lower-luminosity ($L\leq L^{\star}$) counterparts.

\end{abstract}

\begin{keywords}
galaxies: high-redshift - galaxies: evolution - galaxies: formation
\end{keywords}

\section{Introduction}
Improving our understanding of the earliest epochs of galaxy formation and evolution relies
fundamentally on the ability to select clean samples of high-redshift galaxies, free from 
significant low-redshift contamination. Traditionally,
samples of high-redshift galaxies have been selected using one of two
complementary photometric techniques. Firstly, Lyman-alpha emitters
(LAEs) are selected from deep imaging using narrow-band filters
centred on the redshifted  Lyman-alpha emission line (e.g. 
\citealt{Hu1999}). Alternatively, Lyman-break galaxies (LBGs) can be
selected from deep broad-band photometry using the Lyman-break, or
``dropout'', technique pioneered  by 
\cite{Steidel1995}.

Studies of high-redshift galaxies selected using both techniques have
made rapid progress recently, thanks to large-area, red-sensitive,
detectors on ground-based telescopes   and the ultra-deep near-infrared  
imaging now possible with WFC3/IR on-board the Hubble Space Telescope
(HST). As a result,  it is now possible to obtain large, statistical,
samples of LBGs/LAEs in the redshift interval $6<z<7$ from the ground
(e.g. 
\citealt{Yoshida2006,Ouchi2008,Ouchi2010,McLure2009}) 
and upwards of one hundred LBGs have now been identified in
the redshift interval $6.5<z<8.5$ with WFC3/IR (e.g. 
\citealt{Bouwens2010,McLure2010,McLure2011,Finkelstein2010}).  
Indeed, recent results have demonstrated the power of these
techniques with the spectroscopic confirmation of LBG-selected
galaxies within the reionization epoch at $7.00<z<7.20$ (\citealt{Vanzella2011,Pentericci2011,Schenker2011,Ono2011}).
 Although LAEs are simply a sub-set of the LBG population,
due to the different selection techniques employed, the study of these
two high-redshift galaxy populations has proceeded largely
independently. As a consequence, one of the key questions in the study
of high-redshift galaxies is determining how LAEs and LBGs are related
and how they are connected to the galaxy populations identified at
lower redshift.

Over the last fifteen years substantial observational effort has been
invested in studying the population statistics of both LAEs and
LBGs. As a result  significant progress has been made towards an
understanding of the luminosity function and clustering properties of
the LAE and LBG populations from $z=3$ to $z=7$ (e.g. 
\citealt{Bouwens2007,Bouwens2010,Bouwens2011,McLure2009,McLure2010,Ouchi2008,Ouchi2010,Reddy2008}).
 While correlation length estimates for LAEs/LBGs are
largely consistent ($r_{0}\simeq 5$~Mpc) and show little evidence for
evolution, the LAE/LBG luminosity functions appear to show substantial
differential evolution. Most studies now agree that the LBG luminosity
function evolves strongly with redshift, with $L^{\star}$ dimming by a
factor of $\geq 2$ between $z=4$ and $z=6$ (e.g. 
\citealt{Bouwens2007,McLure2009,McLure2010,Yoshida2006}).
 In contrast, the
observed LAE luminosity function does not appear to evolve between
$z=3$ and $z=6$ (e.g. 
\citealt{Shimasaku2006,Ouchi2008}),
although the latest results indicate a moderate $\simeq$ 30\% drop in
$L^{\star}_{Ly\alpha}$ between $z=5.7$ and $z=6.6$ 
\citep{Ouchi2010}.

Studying the Ly$\alpha$ emission properties of LBG-selected samples is a 
powerful method for improving our
understanding of this confusing picture (eg. \citealt{Stark2010,JiangLinhua2011}). Although spectroscopic
follow-up of the LBG population reveals Ly$\alpha$ emission in objects
covering the full span of UV luminosity, the strongest Ly$\alpha$
emission (equivalent width $\gtsim$ 100\AA) is only observed in
fainter objects.  This was first indicated by 
\cite{Shapley2003},
who observed the mean rest-frame Ly$\alpha$ equivalent width (EW) increasing at
fainter magnitudes for $z\sim3$ LBGs, and has since been noted by many
authors over a wide redshift range (eg. 
\citealt{Ando2006,Vanzella2009,Stark2010}).
Furthermore, studies of how the fraction
of strong Ly$\alpha$ emitters evolves as a function of redshift offer
the prospect of providing crucial information on the dust content of
LBGs and, potentially, on the neutral fraction of the IGM. 
Recent work by \cite{Stark2010,Stark2011} has shown that, at fixed UV luminosity, the 
fraction of LBGs showing Ly$\alpha$ emission with EW $\geq 25$\AA~ 
increases by 55\% between $z=4$ and $z=6$. More recently this analysis has been extended to spectroscopically
targeted $z\sim7$ samples 
\citep{Fontana2010,Schenker2011,Pentericci2011,Ono2011}
with the results from different
groups consistently showing a drop in the Ly$\alpha$ fraction at $z>7$, tentatively attributed 
to a change in the neutral fraction in the IGM.

Recent theoretical studies find differing results when trying to explain the observed fractions of Ly$\alpha$ emitting galaxies amongst LBG samples.  \cite{DayalPratika2011} are able to broadly reproduce the observed trend of increasing Ly$\alpha$ emission with decreasing UV luminosity for an EW$>$55\AA~cut at $z\sim6$ from their cosmological SPH simulation of reionization. In agreement with recent observations, 
 \cite{DayalPratika2011} also find a decrease in the fraction of strong Ly$\alpha$ emitters between $6<z<7$, though this is due mainly to changes in dust distributions, not a change in the neutral fraction of the IGM. However, they also conclude that all LBGs at $z\sim6$ should display Ly$\alpha$ in emission with EW$>$20\AA, in disagreement with the observations of \cite{Stark2011}. 
\cite{Forero-RomeroJ.E.2011} are able to reproduce very well the observed fraction of EW$>$55\AA~ Ly$\alpha$ emitters in the $4.5<z<6.0$ sample from \cite{Stark2010}, by requiring the escape fraction of Ly$\alpha$ photons to decrease with increasing UV-luminosity.  
In addition, they are are also able to reproduce the observed drop in EW$>$25\AA{ } Ly$\alpha$ emitters above $z>6.3$ observed by \cite{Schenker2011}, by fitting to the observed evolution in the LAE luminosity function at these redshifts.  
Neither of these studies allow for peculiar velocities (outflows/inflows) within the ISM which \cite{Dijkstra2011} find to produce a large effect on the detectability of Ly$\alpha$, with outflows from the galaxies boosting the observed EWs.
 
Despite the large amounts of effort invested in spectroscopically
observing the properties of LBGs over recent years,  one area of
parameter space which has been relatively unexplored is that occupied by luminous LBGs
at $z\geq 6$. The relatively poor statistics in this
redshift-luminosity r\'{e}gime are simply a reflection of the relative
rarity of $L\geq L^{\star}$ galaxies at $z\geq 6$, combined with the
small areas typically covered by survey fields with the deep,
multi-wavelength, imaging necessary to reliably select such objects. In this
paper we address this issue by presenting deep, red sensitive, optical
spectroscopy of luminous ($L\geq 2L^{\star}$) LBGs in the
redshift interval $6.0<z<6.5$,  photometrically selected from an area
of 0.25 sq. degrees within the UKIDSS 
\citep{Lawrence2007}
Ultra-Deep Survey (UDS). Armed with deep spectroscopic observations,
the combination of large area and deep optical/near-infrared imaging available
in the UDS allows us to investigate three important issues. Firstly,
we are able to investigate whether or not the photometrically selected
LBG samples used to constrain the evolution of the bright end of the
UV-selected galaxy luminosity function (e.g. 
\citealt{McLure2009}) are
significantly contaminated by low-redshift interlopers and/or active
galactic nuclei (AGN). Secondly, we are able to place the best
available constraints on the fraction of strong Ly$\alpha$ emitters
amongst luminous LBGs at $z\geq 6$. Finally, we are able to
investigate whether previous reports of a very low fraction of strong
Ly$\alpha$ emitters amongst samples of luminous $z\simeq 6$ LBGs have been 
biased due to contamination by low-redshift interlopers or small number statistics.

The structure of the paper is as follows.  In Section 2 we describe
the  initial sample selection and spectroscopic observations before
proceeding to present the spectra themselves. In Section 3 we
describe our technique for accurately measuring the Ly$\alpha$
equivalent widths (EWs) and investigate the star-formation rates (SFRs) and Ly$\alpha$ photon escape fraction 
of our LBG sample, before estimating the fraction of
luminous $z\simeq 6$ LBGs which are strong Ly$\alpha$ emitters.  In
Section 4 we compare our results to those from the recent literature
and discuss the potential uncertainties and biases associated with
estimating the Ly$\alpha$ emitter fraction at high redshift. In
Section 5 we present our conclusions.  Throughout the paper we assume
a cosmology with $H_0=70$ km s$^{-1}$Mpc$^{-1}$, $\Omega_m=0.3$,
$\Omega_{\Lambda}=0.7$.  All magnitudes are quoted in the AB system (\citealt{Oke1983}).

\begin{figure*}
  \centering
  \includegraphics[width=6in, trim=1.5cm 13cm 3cm 6cm,clip]{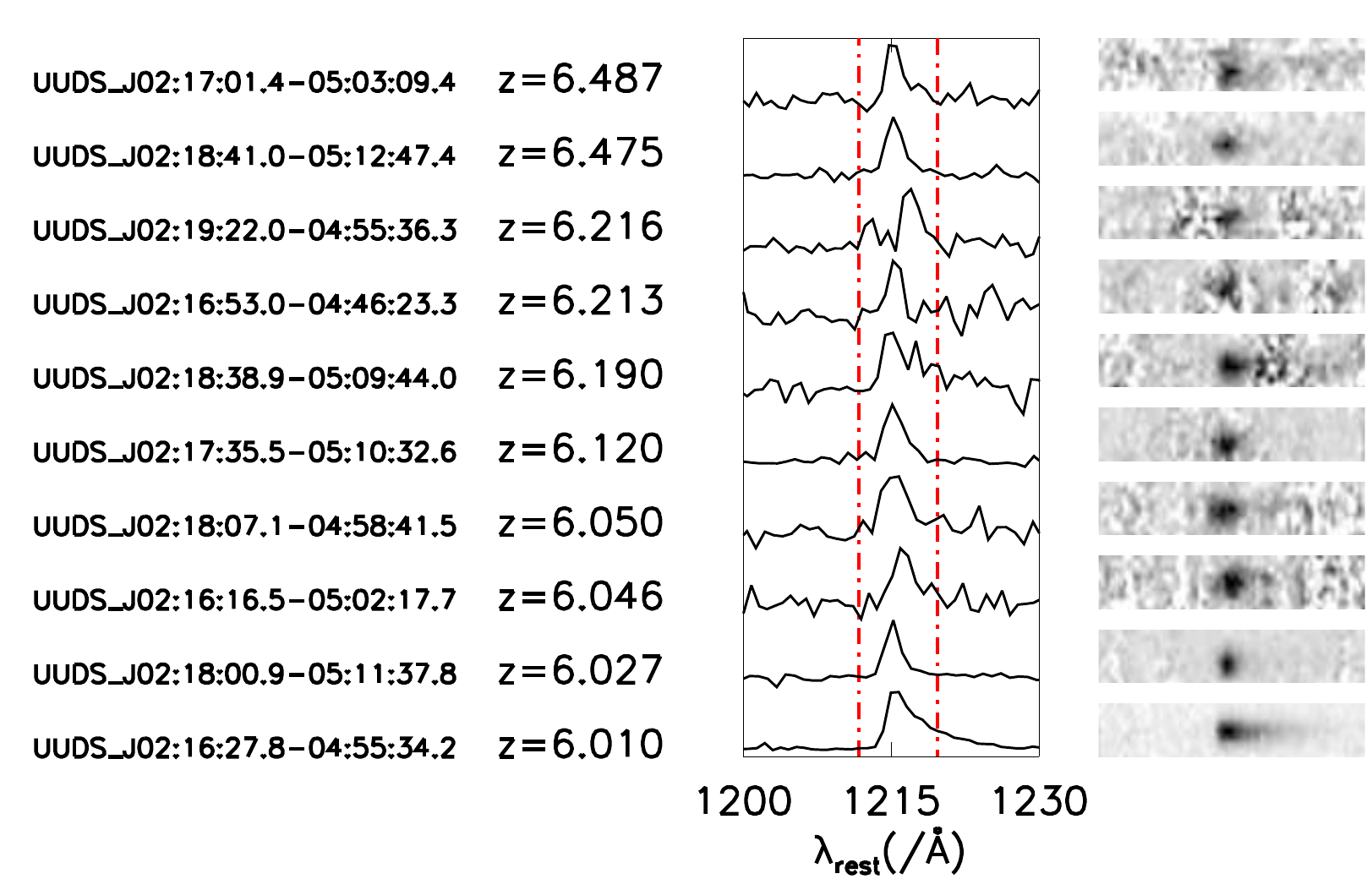}
\caption{Spectra of the confirmed $z>6$ objects showing the one-dimensional (1D) spectra to the left and the corresponding two-dimensional (2D) spectra on the right.  All 1D spectra are plotted on the same flux scale, except for the spectrum of the AGN (UUDS$\_$J021627.8) which has been scaled by a factor of 0.5.}
\label{spectra}
\end{figure*}

\section{Imaging data, sample selection and spectroscopy}
The sample of $z\geq 6$ LBG candidates targeted for spectroscopic
follow-up is a sub-set of that originally selected by McLure et
al. (2009), using photometric redshift analysis, to investigate the bright-end of the $z=6$ galaxy
luminosity function. Although a full description of the original
selection process is provided in McLure et al. (2009), in this section
we briefly review the most relevant details before proceeding to
describe the spectroscopic observations.

\subsection{Imaging data and LBG selection}
The parent sample of $z\geq 6$ LBG candidates was originally
selected via a photometric redshift  analysis, which exploited the
deep optical and near-infrared imaging available within the area covered by
the UKIDSS Ultra-Deep Survey (UDS). The UDS is the deepest of the five
near-infrared surveys being undertaken at the UK infra-red telescope (UKIRT)
which together comprise the UK infra-red Deep Sky Surveys (UKIDSS;
Lawrence et al. 2007). 

The latest ESO-public data-release for the UDS
(DR8) features near-infrared imaging over an area of 0.8 square degrees to
$5\sigma$-depths of $J=24.9, H=24.2\,\, \&\,\, K=24.7$ (2\asec diameter apertures).  
Within the UDS field, complimentary deep Subaru
optical imaging is available from the  Subaru/XMM Deep Survey (SXDS),
which provides optical imaging to $5\sigma$ depths (2\asec diameter
apertures) of $B=27.9, V=27.3, R=27.2, i'=27.2$ \& $z'=26.1$ (Furusawa
et al. 2008). 

The useful overlap region between the UDS near-infrared and SXDS
optical imaging covers an area of $\simeq 0.65$ square degrees, and it
was from this overlap region that McLure et al. (2009) used a
photometric redshift analysis to select luminous ($z'<26$)
 LBG candidates in the redshift interval
$4.5<z<6.5$. The original selection utilised the full information 
available from the SXDS optical bands and the UKIDSS DR1 $J-$ and $K-$band 
near-infrared imaging.  Our updated photometric redshift analysis, using the 
UKIDSS DR8 $J$, $H$ and $K$-band imaging, is described in greater detail in 
Section 2.2.3.  The parent sample for spectroscopic follow-up was
comprised of those candidates with a high probability of lying at
$z\geq 6$ based on the redshift probability density function 
returned by the photometric redshift analysis.

  \addtocounter{footnote}{1}

\begin{table*}
 \centering
 \begin{minipage}{160mm}
  \caption{The spectroscopic sample of $z\geq6$ UDS LBGs.  Column 1 lists the UDS object IDs (incorporating the J2000 coordinates).  
Column 2 gives the spectroscopic redshift and errors derived from the FORS2 spectra, taking the spectroscopic redshift from the peak of the Ly$\alpha$ line.  Columns 3 \& 4 give the quality of the 
spectroscopic redshift (A-C, see text for details) and the photometric redshift derived in Section 2.3 respectively. The remaining columns list the $z^{\prime}$-band and $z_{921}$-NB observed magnitudes (corrected to total) plus errors, absolute UV magnitude derived from the $z_{921}$-NB photometry and the measured Ly$\alpha$ fluxes and EWs (derived from spectra, see text for more details).}
  \scalebox{0.75}{
  \begin{tabular}{@{}lcccccccc@{}}
  \hline
  \hline
   ID & $z_{spec}$ & Quality & $z_{phot}$ & m$_z$ &  m$_{912}$ & M$_{UV}$  & Ly$\alpha$ flux & Ly$\alpha$ EW\\
                  &            &       &     &  &   & & (/$10 ^{-18}$ ergs s$^{-1}$ cm$^{-2}$) & (/\AA)\\
 \hline

  UUDS\_J021800.90-051137.8 & 6.027 $\pm$ 0.002 &  A & $6.0^{+0.1}_{-0.1}$ & 25.36 $\pm$ 0.08 & 25.30 $\pm$ 0.09 & $-$21.40 & \phantom{0}44.9 $\pm$ 7.4\phantom{0} & \phantom{0}54.7 $\pm$ 11.3\phantom{0} \\
  UUDS\_J021616.53-050217.7 & 6.046 $\pm$ 0.004 &  A & $5.9^{+0.2}_{-0.1}$ & 25.35 $\pm$ 0.15 & 25.84 $\pm$ 0.13 & $-$20.86 & \phantom{0}17.8 $\pm$ 3.1$^{\mathit{a}}$ & \phantom{0}36.8 $\pm$ 8.5\phantom{00}\\
  UUDS\_J021807.14-045841.5 & 6.050 $\pm$ 0.003 &  A & $6.0_{-0.1}^{+0.2}$ & 25.02 $\pm$ 0.12 & 24.99 $\pm$ 0.10 & $-$21.72 & \phantom{0}29.4 $\pm$ 2.6\phantom{0} & \phantom{0}27.6 $\pm$ 4.1\phantom{00} \\
  UUDS\_J021816.33-051116.6 & 6.114 $\pm$ 0.009 &  C & $6.1^{+0.3}_{-0.3}$ & 25.55 $\pm$ 0.15 & 25.42 $\pm$ 0.14 & $-$21.29 & \phantom{00}7.3 $\pm$ 1.7\phantom{0} & \phantom{0}11.6 $\pm$ 3.1\phantom{00}\\
  UUDS\_J021735.34-051032.6 & 6.120 $\pm$ 0.003 &  A & $6.0^{+0.1}_{-0.1}$ & 25.21 $\pm$ 0.12 & 25.47 $\pm$ 0.15 & $-$21.25 & \phantom{0}31.4 $\pm$ 2.6\phantom{0} & \phantom{0}46.8 $\pm$ 8.4\phantom{00}  \\
  UUDS\_J021838.90-050944.0 & 6.190 $\pm$ 0.014 &  A & $6.0_{-0.1}^{+0.2}$ & 25.21 $\pm$ 0.11 & 25.13 $\pm$ 0.11 & $-$21.62 & \phantom{0}36.4 $\pm$ 3.4\phantom{0} & \phantom{0}39.4 $\pm$ 6.0\phantom{00} \\
  UUDS\_J021653.00-044623.3 & 6.213 $\pm$ 0.004 &  A & $6.5_{-0.4}^{+0.4}$ & 25.75 $\pm$ 0.21 & 25.34 $\pm$ 0.09 & $-$21.41 & \phantom{00}9.5 $\pm$ 1.1\phantom{0} & \phantom{0}12.5 $\pm$ 3.1\phantom{00}\\
  UUDS\_J021922.01-045536.3 & 6.216 $\pm$ 0.009 &  A & $6.4_{-0.8}^{+0.8}$ & 25.79 $\pm$ 0.22 & 26.25 $\pm$ 0.20 & $-$20.50 & \phantom{0}16.8 $\pm$ 2.2\phantom{0} & \phantom{0}50.7  $\pm$ 12.8\phantom{0}\\
  UUDS\_J021841.02-051247.4 & 6.475 $\pm$ 0.003 &  A & $5.9_{-0.2}^{+0.4}$ & 25.80 $\pm$ 0.23 & 25.46 $\pm$ 0.16 & $-$21.36 & \phantom{0}18.2 $\pm$ 3.1\footnote{Ly$\alpha$ flux calibration tied to the $z-$band photometry using an average correction factor, see text for details.} & 27.8 $\pm$ 6.8$^{\mathit{b}}$ \\
  UUDS\_J021701.44-050309.4 & 6.487 $\pm$ 0.004 & A & $6.0_{-0.3}^{+0.3}$ & 25.91 $\pm$ 0.19 & 25.03 $\pm$ 0.10 & $-$21.79 & \phantom{0}10.2 $\pm$ 1.8$^{\mathit{a}}$ & \phantom{0}10.6 $\pm$ 2.3\footnote{EW values potentially affected by Ly$\alpha$ contributing to the NB photometry used to derive the UV continuum value.}\phantom{0} \\
\hline
\hline
\end{tabular}}
\label{tab:UDSlya}
\end{minipage}
\end{table*}

\addtocounter{footnote}{1}

\subsection{Spectroscopy}
The spectroscopic data analysed in this study were obtained between
October 2007 and January 2011 with the FORS2 spectrograph on the VLT
as part of the systematic spectroscopic follow-up of the UDS obtained
through the ESO large programme ESO 180.A-0776 (UDSz;
P.I. O. Almaini). Given that full details of UDSz will be presented in
Almaini et al. (2012, in preparation), only the most relevant details are
provided here.

The UDSz programme was allocated a total of 235 hours of observations,
with 93 hours allocated for observations with the VIMOS spectrograph
and 142 hours allocated for FORS2 observations. The primary science
driver for UDSz was to obtain spectroscopic observations of a
representative sample of $K-$band selected galaxies ($K<23$)
photometrically pre-selected to lie at redshift $z_{phot}\geq
1.0$. These spectroscopic observations were designed to support the
primary science driver of the UDS survey, which is to study the
assembly and evolution of massive galaxies at $z\geq 1$. Within this
context, the large allocation of FORS2 time (20 MXU masks, 5 hours of
integration each) was designed to provide sensitive, red optical,
spectra of massive galaxies in the  redshift interval $1.0<z<1.5$.

Each of the 20 FORS2 masks covers an area of 6.8$^{\prime}$ $\times$
6.8$^{\prime}$ and typically allowed for the allocation of 30-35
science slits (based on a slit length of 8\asec). During the mask
design process targets from the core sample of $K<23$ galaxies were
allocated with the highest priority. However, in order to fill each
mask, several other samples of additional science targets were also
allocated slits on a best-effort basis. The list of high-redshift LBG
candidates from McLure et al. (2009) with a high probability of lying
at $z\geq 6$ was included as one of the samples of additional science
targets. A total of fourteen $z\geq 6$ LBG candidates were observed as
part of UDSz, each receiving a total of 5-hours of on-source
integration with  the GRS$\_$300I grism
($6000$\AA$<\lambda<10000$\AA\,  with $R=660$). Data reduction was
performed using a modified version of the FORS2 pipeline,  details of
which are provided in Pearce et al. (2011, in preparation).

\subsubsection{Spectra of confirmed $z\geq6$ sources}
Of the fourteen galaxies for which spectra were obtained we are able to assign secure redshifts to 
eleven sources according to Ly$\alpha$ line detections, placing them at $z>6$.  The 1D spectra for the
spectroscopically confirmed $z>6$ Ly$\alpha$ emitters are displayed
in Figure~\ref{spectra} and the measured line fluxes and EWs are reported
in Table \ref{tab:UDSlya}. Many of the spectra show
continuum red-wards of the emission line as well as the distinctive
asymmetrical line profile indicative of Ly$\alpha$.  Each redshift is assigned a quality flag,
defined as in \cite{Vanzella2009}, with values of A (unambiguous
detection), B (likely detection) and C (uncertain detection). It should be noted that the redshifts
for the 11 confirmed $z>6$ sources are measured from the position of peak flux
within the Ly$\alpha$ line, and are hence susceptible to shifts in the
peak position due any velocity structure of the neutral hydrogen
within the galaxy or the surrounding IGM.  The average velocity shift measured for Ly$\alpha$ in lower redshift samples is of order +350 km s$^{-1}$ \citep{Adelberger2003,Shapley2003}, since the Ly$\alpha$ photons are more likely to escape the galaxy when they are scattered off neutral Hydrogen that is travelling away from us, giving the photons a Doppler shift that takes them off resonance.

One of our targets (UUDS\_J021627.80-045534.2) is a low-luminosity active 
galactic nucleus (AGN) at a redshift of $z=6.01$ (Ly$\alpha$ FWHM $\simeq 1600$ km s$^{-1}$). This 
object was initially identified as a potential ultra-luminous $z\simeq 6$ LBG candidate by 
McLure et al. (2006) and was subsequently targeted as a high-redshift quasar candidate 
by Willott et al. (2009). In Willott et al. (2009) this source is referred to as 
CFHQS J021627$-$045534, and was confirmed as a low-luminosity quasar at $z=6.01$ based on a 
four-hour spectrum obtained with the GMOS spectrograph on Gemini. 
However, it is important to note that CFHQS J021627$-$045534 was the only $z\simeq 6$ AGN identified by Willott et al. (2010) over
a search area of 4.47 sq. degrees, down to a limit of $z^{\prime}=24.5$. Based on this fact, and the steepness of the $z\simeq 6$ UV-selected galaxy
luminosity function, it seems reasonable to conclude that the luminous, type I AGN contamination of $26\geq z^{\prime}\geq 25$ LBG samples at $z\simeq 6$ is negligible.

\begin{figure*}
  \center
  \subfigure{\includegraphics[width=2.2in,trim=3.8cm 13.2cm 1.5cm 4.5cm,clip]{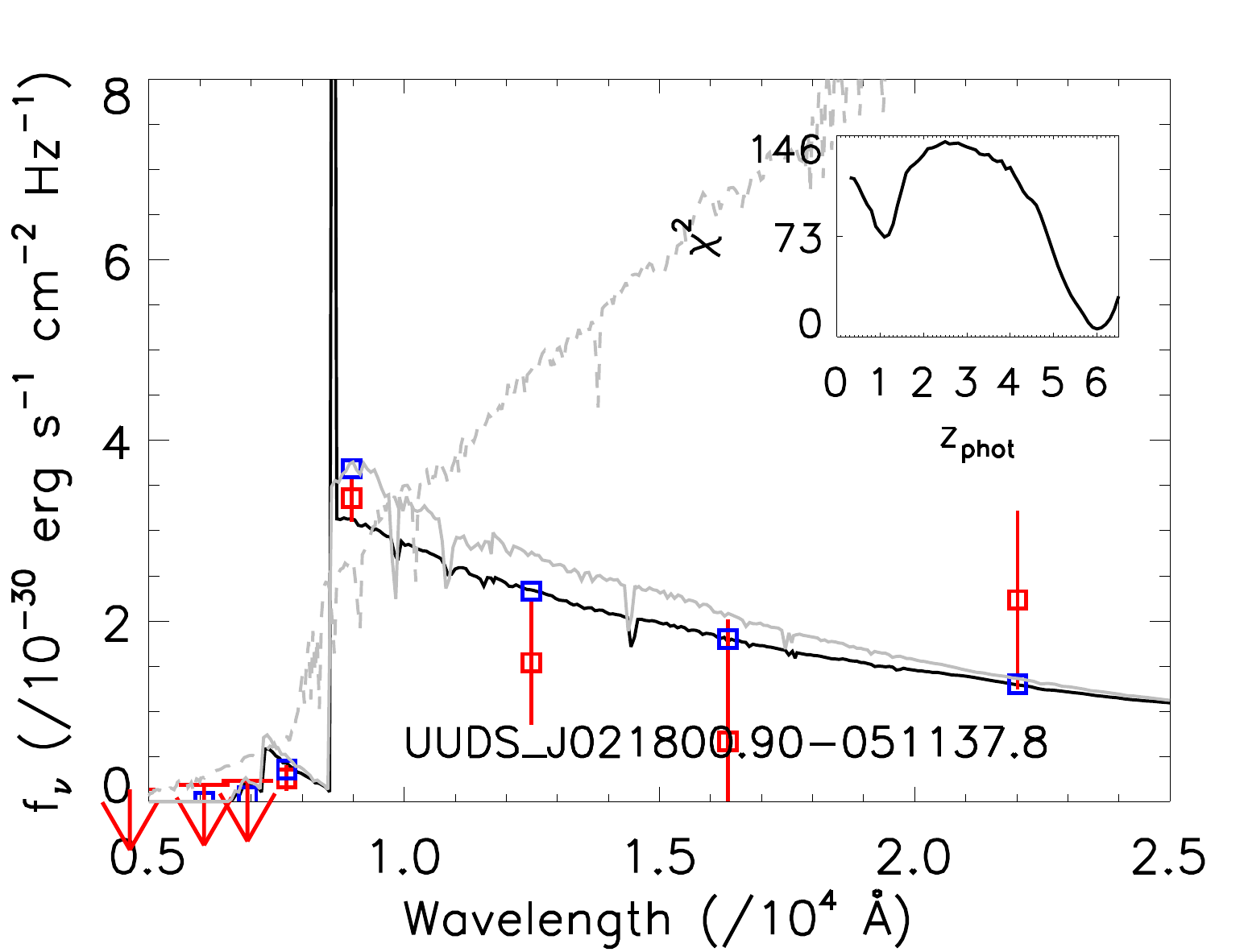}}
  \subfigure{\includegraphics[width=2.2in,trim=3.8cm 13.2cm 1.5cm 4.5cm,clip]{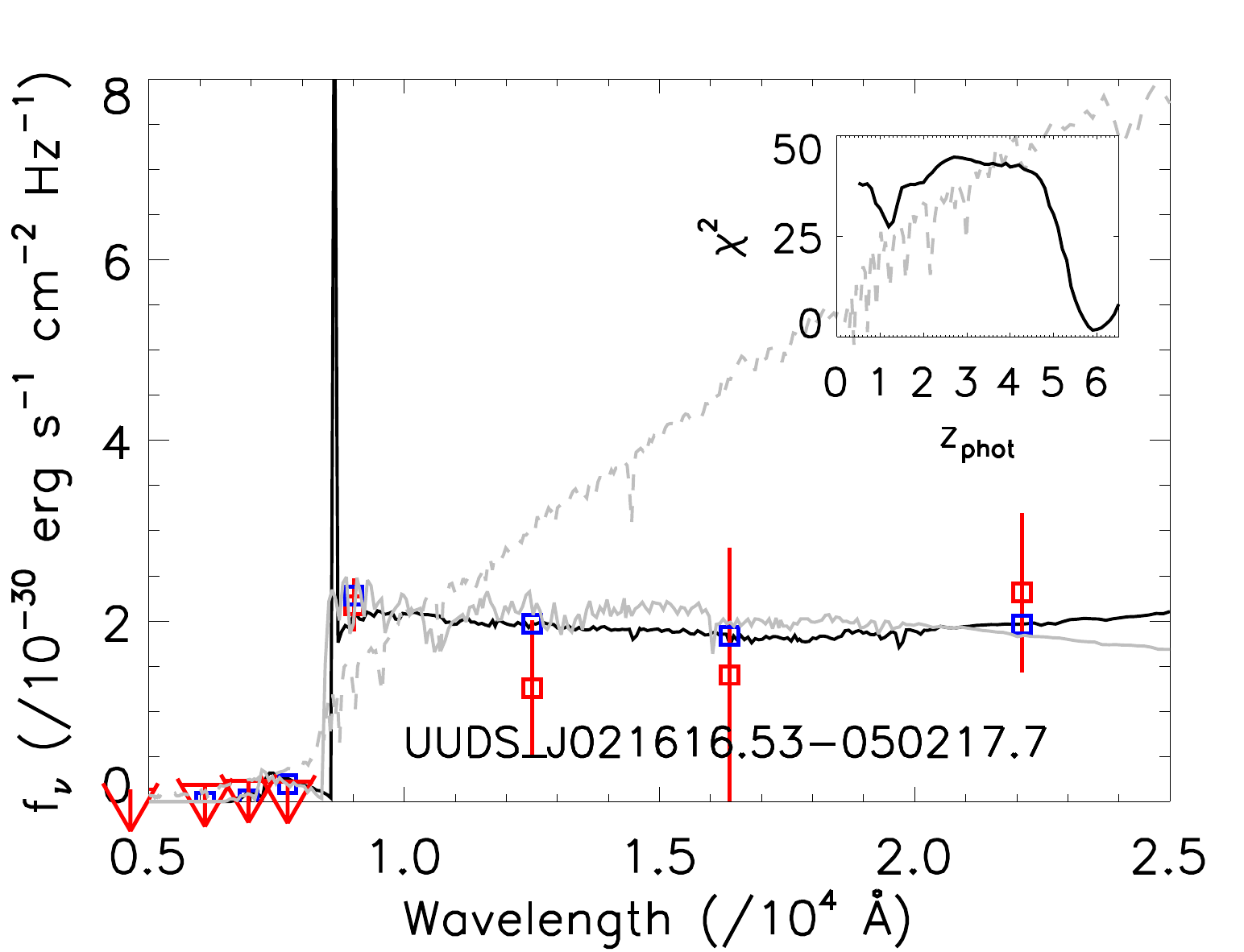}}
  \subfigure{\includegraphics[width=2.2in,trim=3.8cm 13.2cm 1.5cm 4.5cm,clip]{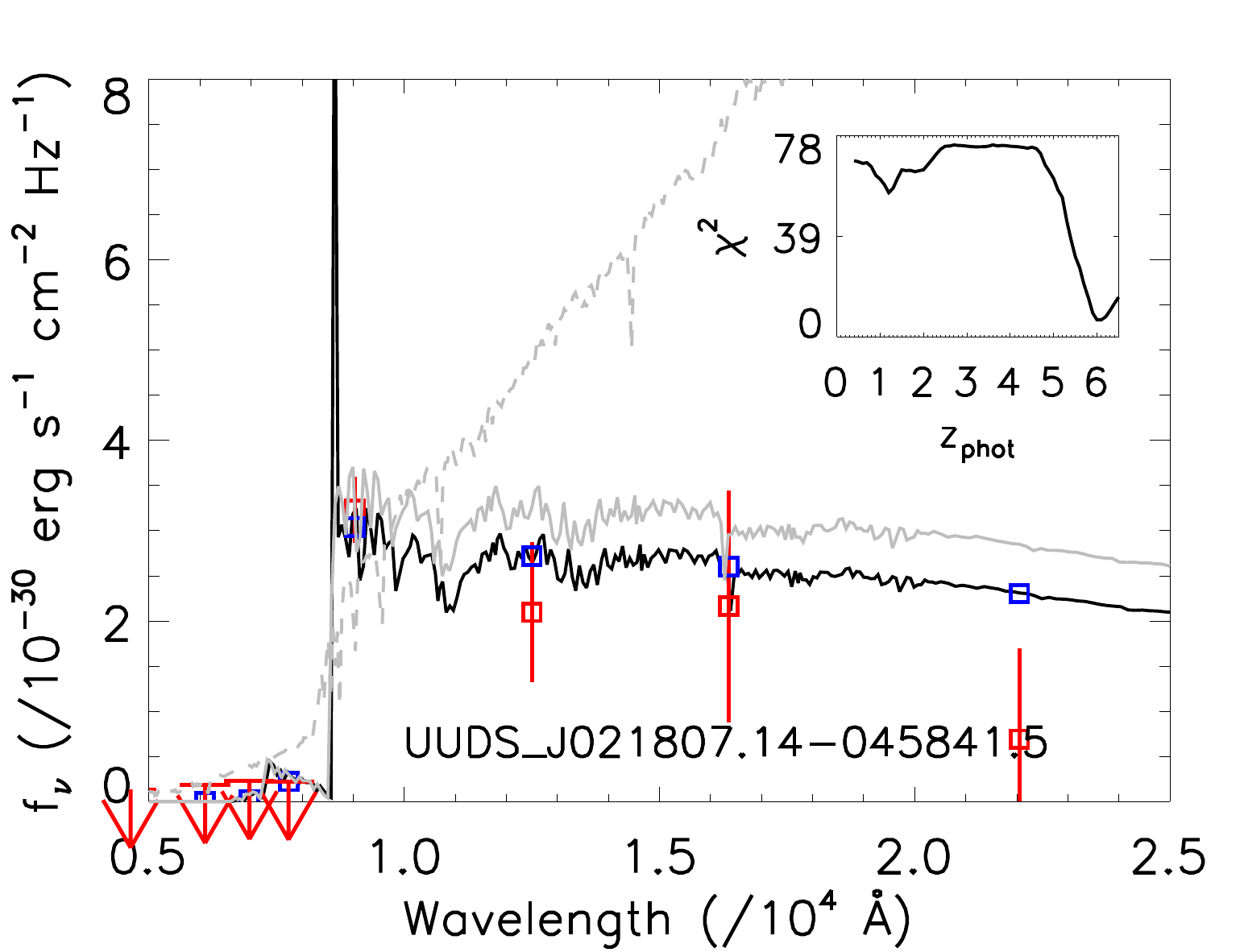}}
  \subfigure{\includegraphics[width=2.2in,trim=3.8cm 13.2cm 1.5cm 4.5cm,clip]{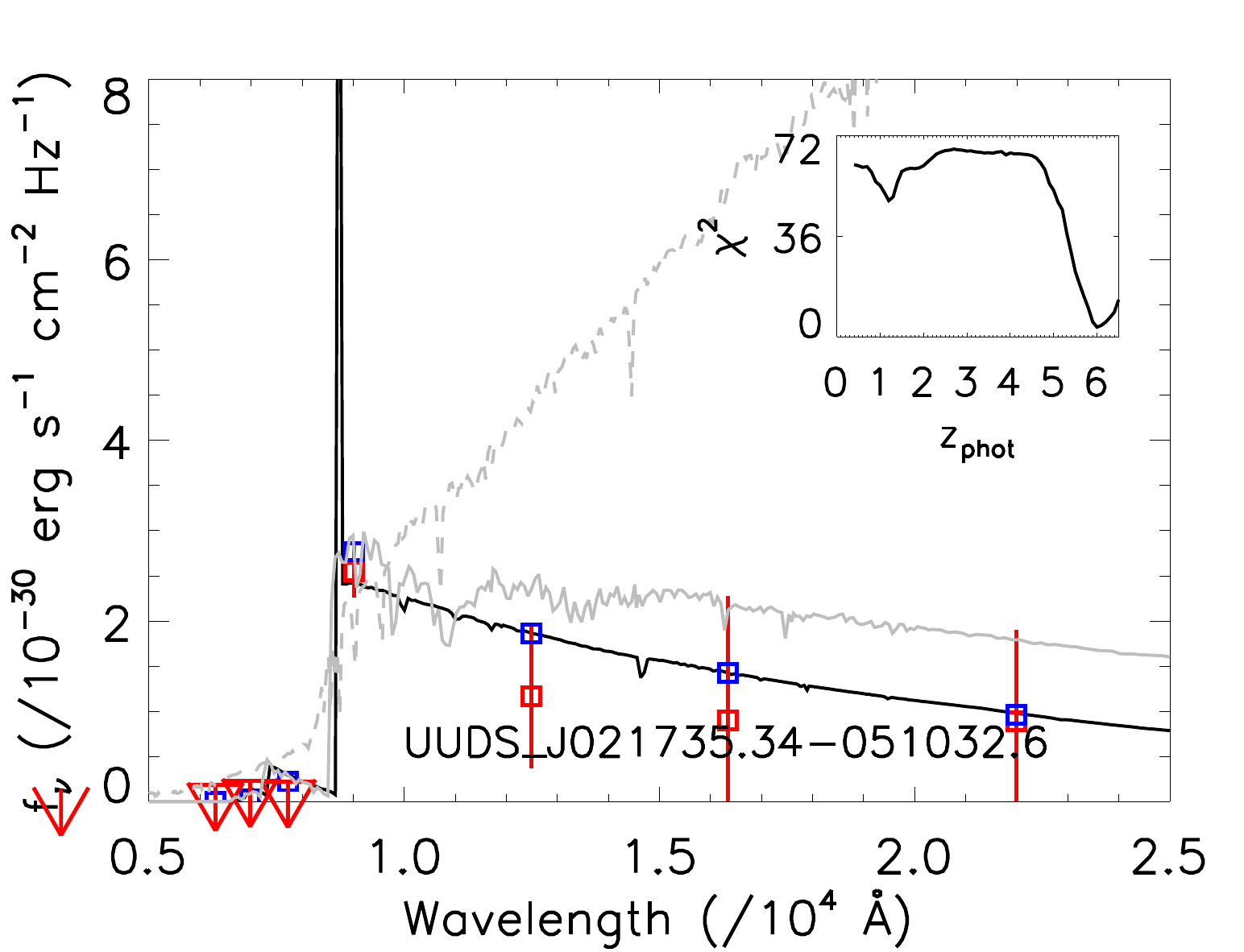}}
  \subfigure{\includegraphics[width=2.2in,trim=3.8cm 13.2cm 1.5cm 4.5cm,clip]{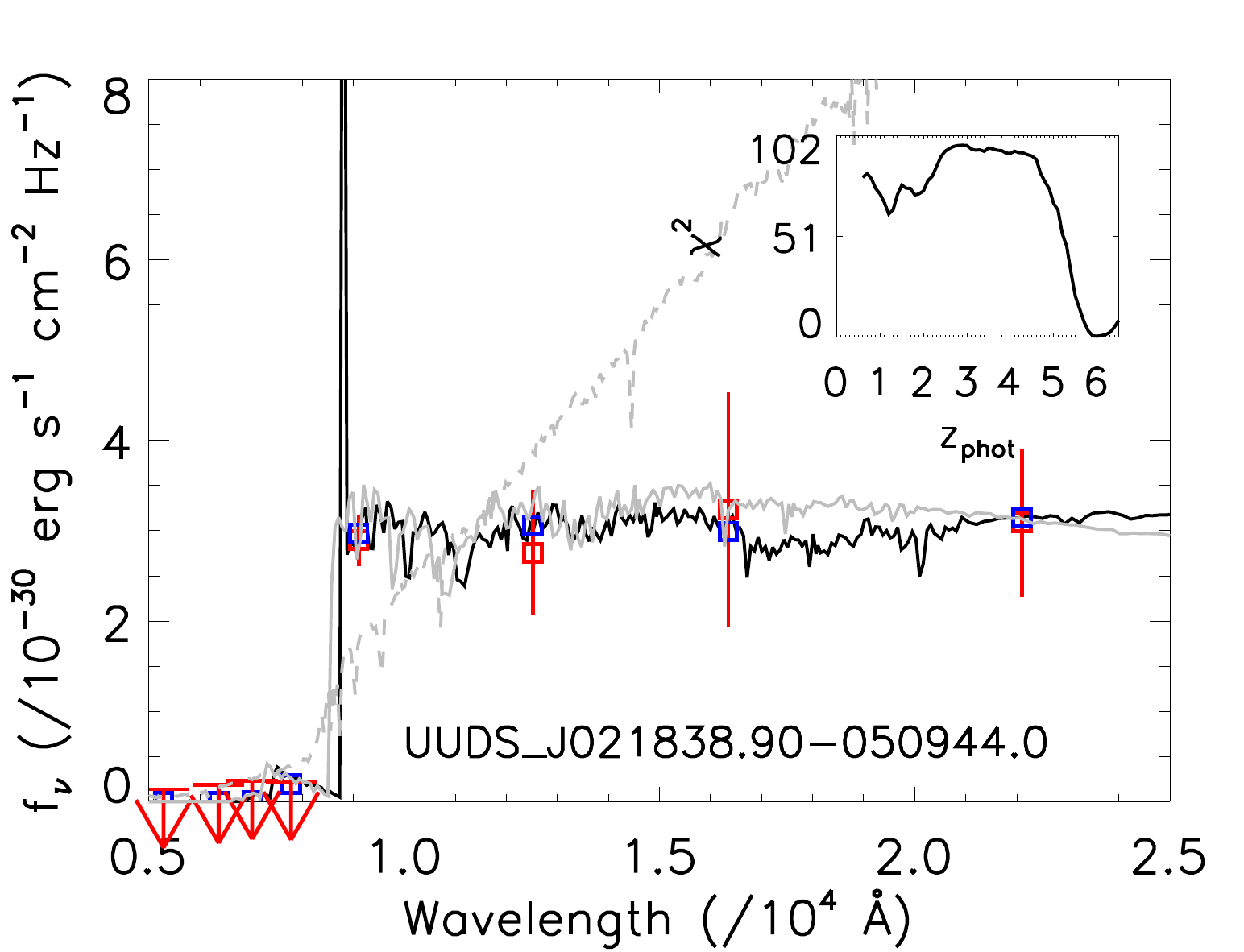}}
  \subfigure{\includegraphics[width=2.2in,trim=3.8cm 13.2cm 1.5cm 4.5cm,clip]{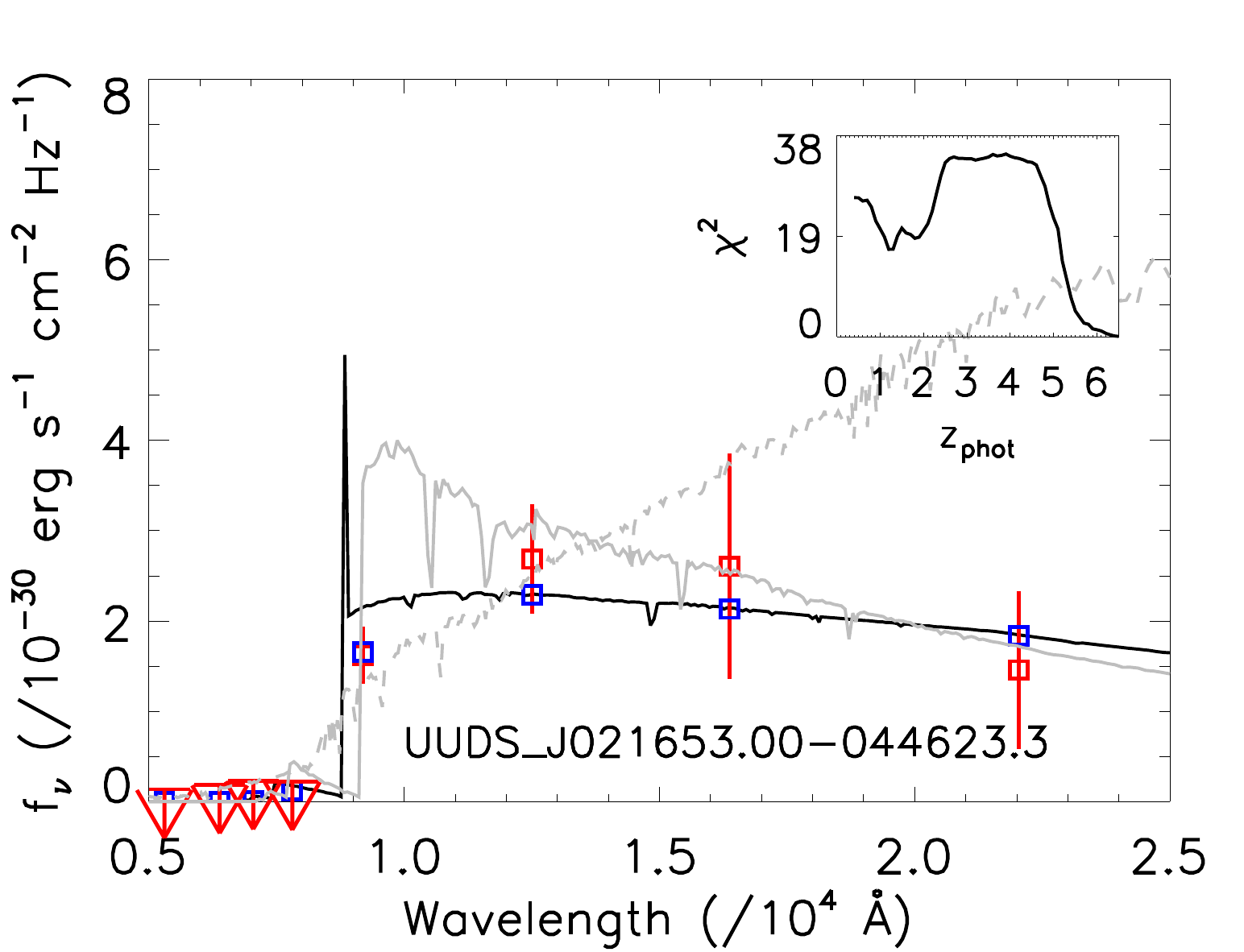}}
  \subfigure{\includegraphics[width=2.2in,trim=3.8cm 13.2cm 1.5cm 4.5cm,clip]{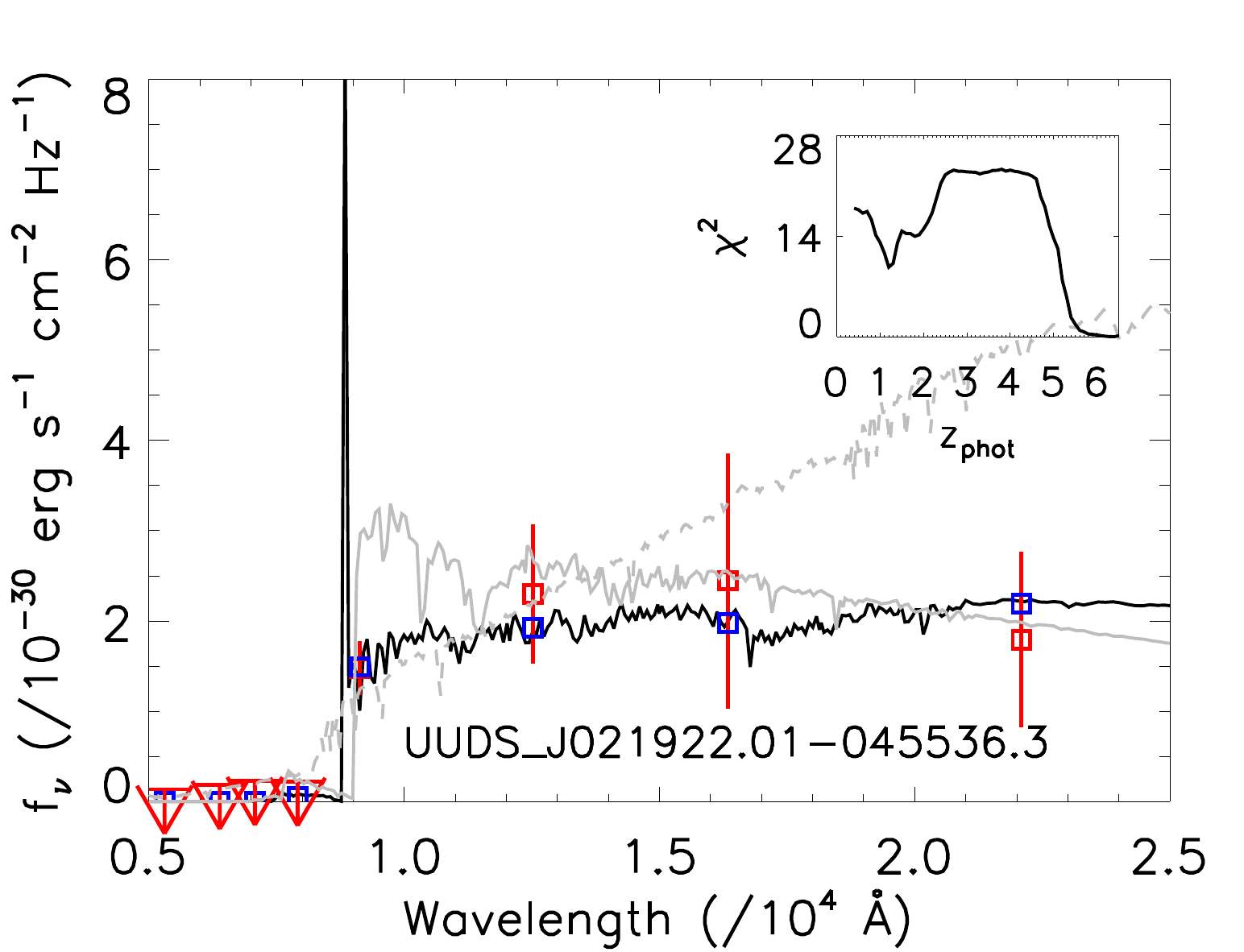}}
  \subfigure{\includegraphics[width=2.2in,trim=3.8cm 13.2cm 1.5cm 4.5cm,clip]{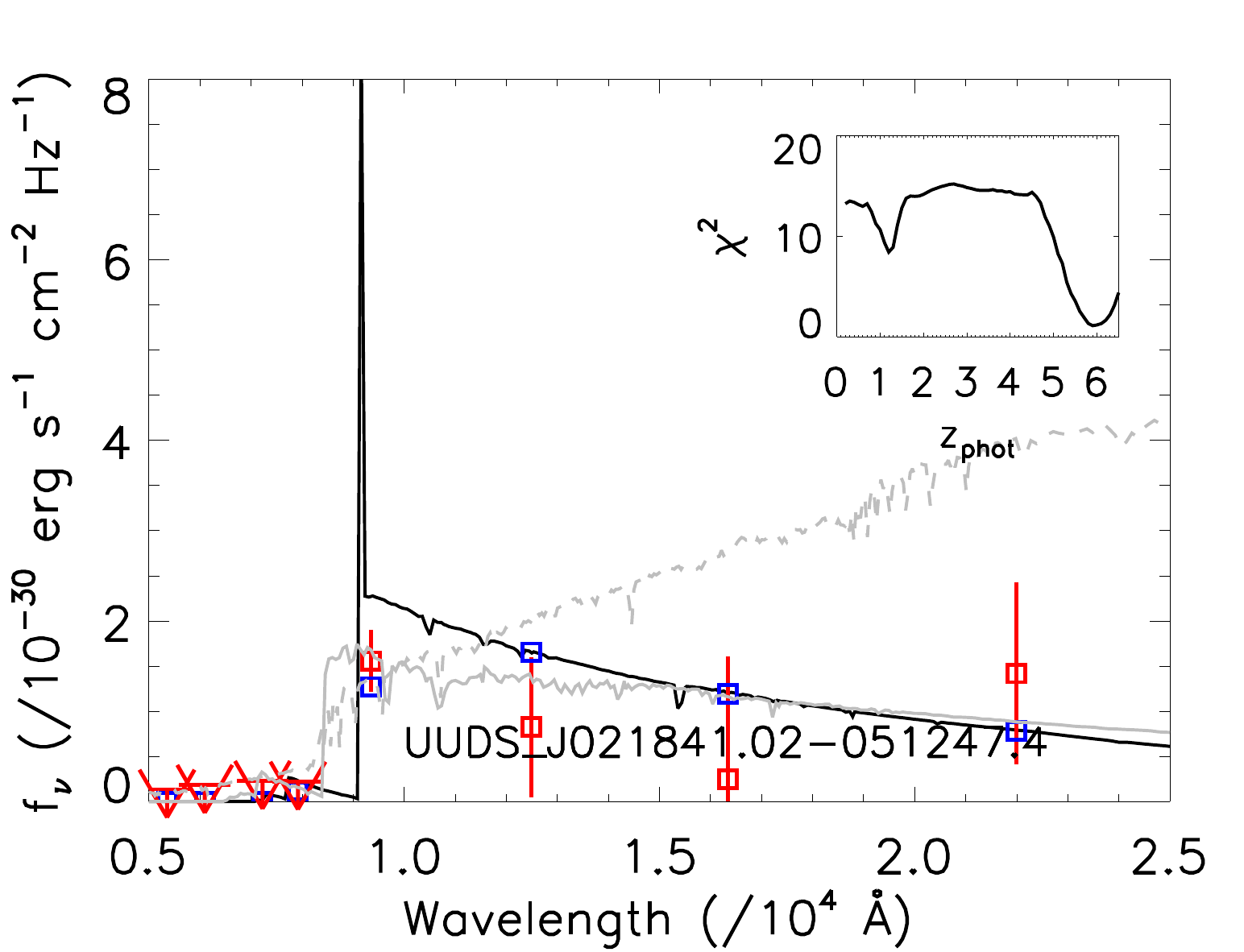}}
  \subfigure{\includegraphics[width=2.2in,trim=3.8cm 13.2cm 1.5cm 4.5cm,clip]{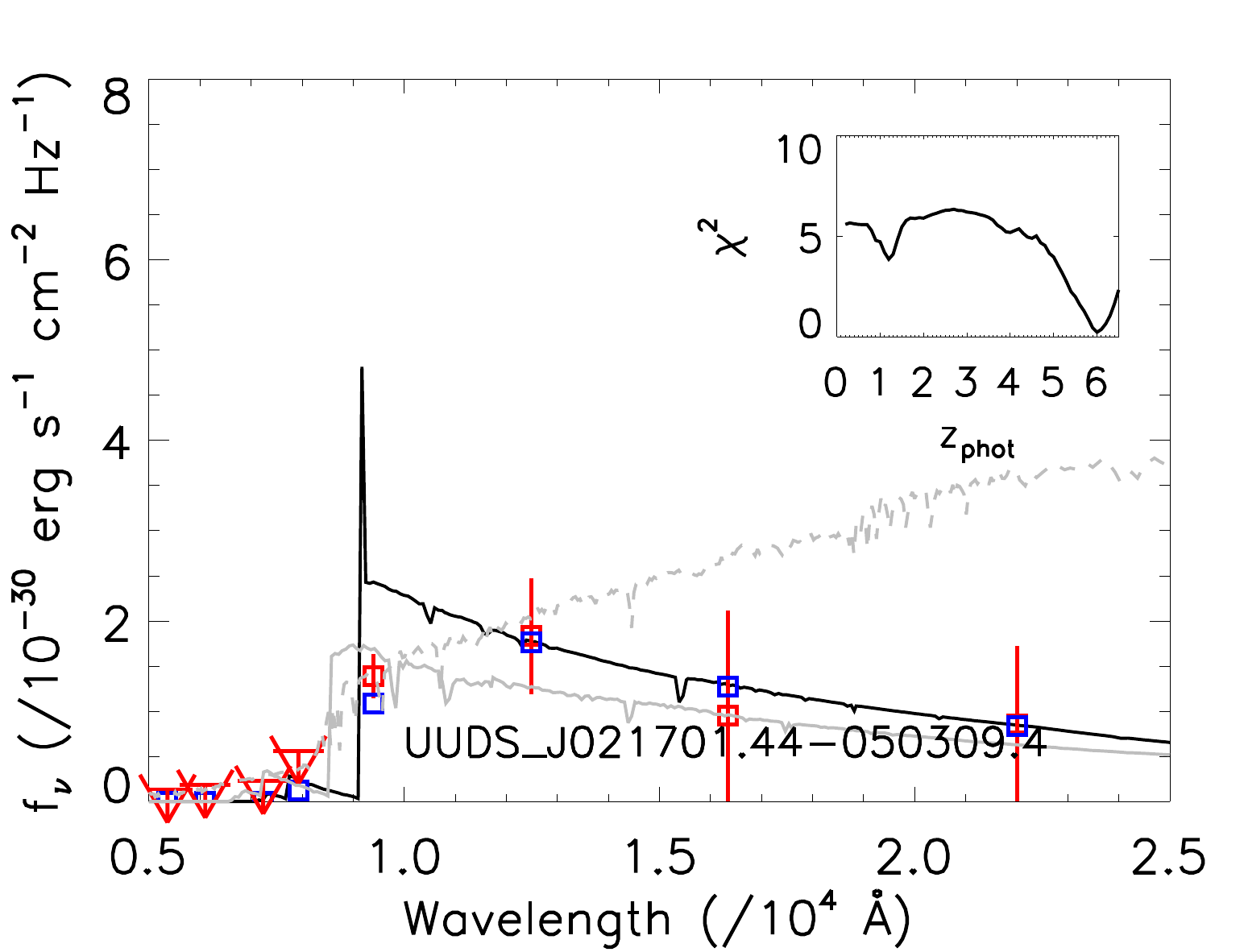}}
  \caption{SED fits to the spectroscopically confirmed $z\geq6$ LBGs showing the $BVRizJHK$ photometry (red data points with error bars), the 
best fitting template with a fixed spectroscopic redshift and including the measured Ly$\alpha$ contribution (black line and blue data points) 
and the best-fitting template assuming no Ly$\alpha$ flux and leaving redshift as a free parameter (grey lines). The best-fitting SED templates at low redshift ($z_{phot}\leq 2$) are plotted as dashed grey lines. The inset panels show the distribution of $\chi^2$ versus redshift.}
\label{fig:photoz}
\end{figure*}

Based on the optical spectra alone, we cannot entirely rule out the possibility that the 
observed Ly$\alpha$ emission could have some contribution from a narrow-line AGN component. However, we 
note that a recent study of UV-selected AGN at $z\sim2-3$ by \cite{Hainline2011} indicates that 
narrow-line AGN are on average much redder than non-AGN LBGs ($\beta=-0.314$ compared to $\beta=-1.49$) 
and show the NV line at $\lambda_{rest}=1240$\AA{ }in emission. Given that our sample have broad-band SEDs which 
are not well reproduced by a standard Type-2 AGN template (see Figure~\ref{fig:photoz}), are not 
detected at 24$\mu$m and do not display NV emission lines in their FORS2 spectra, we conclude 
that the UV continuum in these objects is dominated by starlight.

\begin{figure*}
  \center
  \subfigure{\includegraphics[width=2.5in,trim=2.8cm 13.2cm 0.5cm 4.5cm,clip]{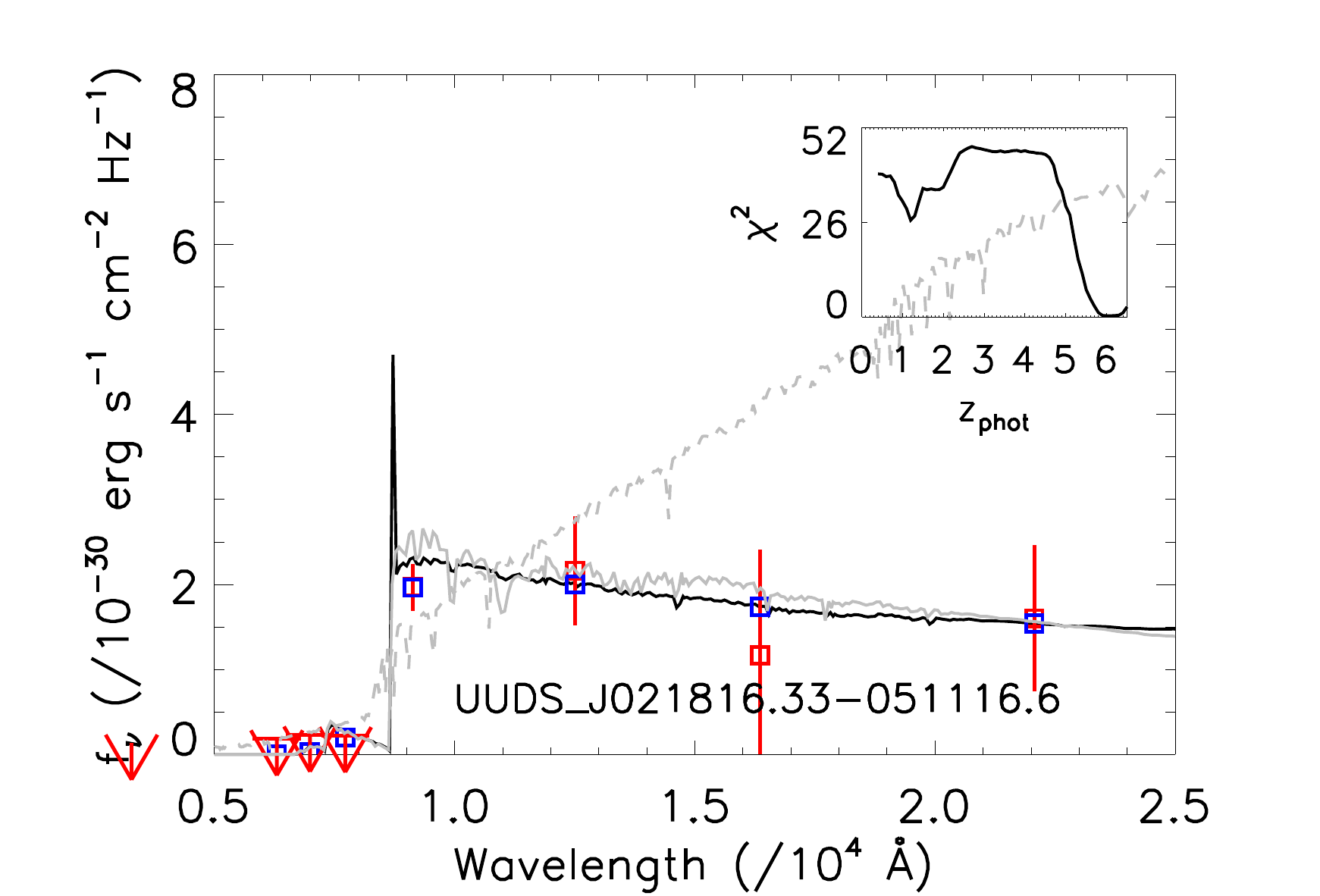}}
  \subfigure{\includegraphics[width=2.5in,trim=2.8cm 13.2cm 0.5cm 4.5cm,clip]{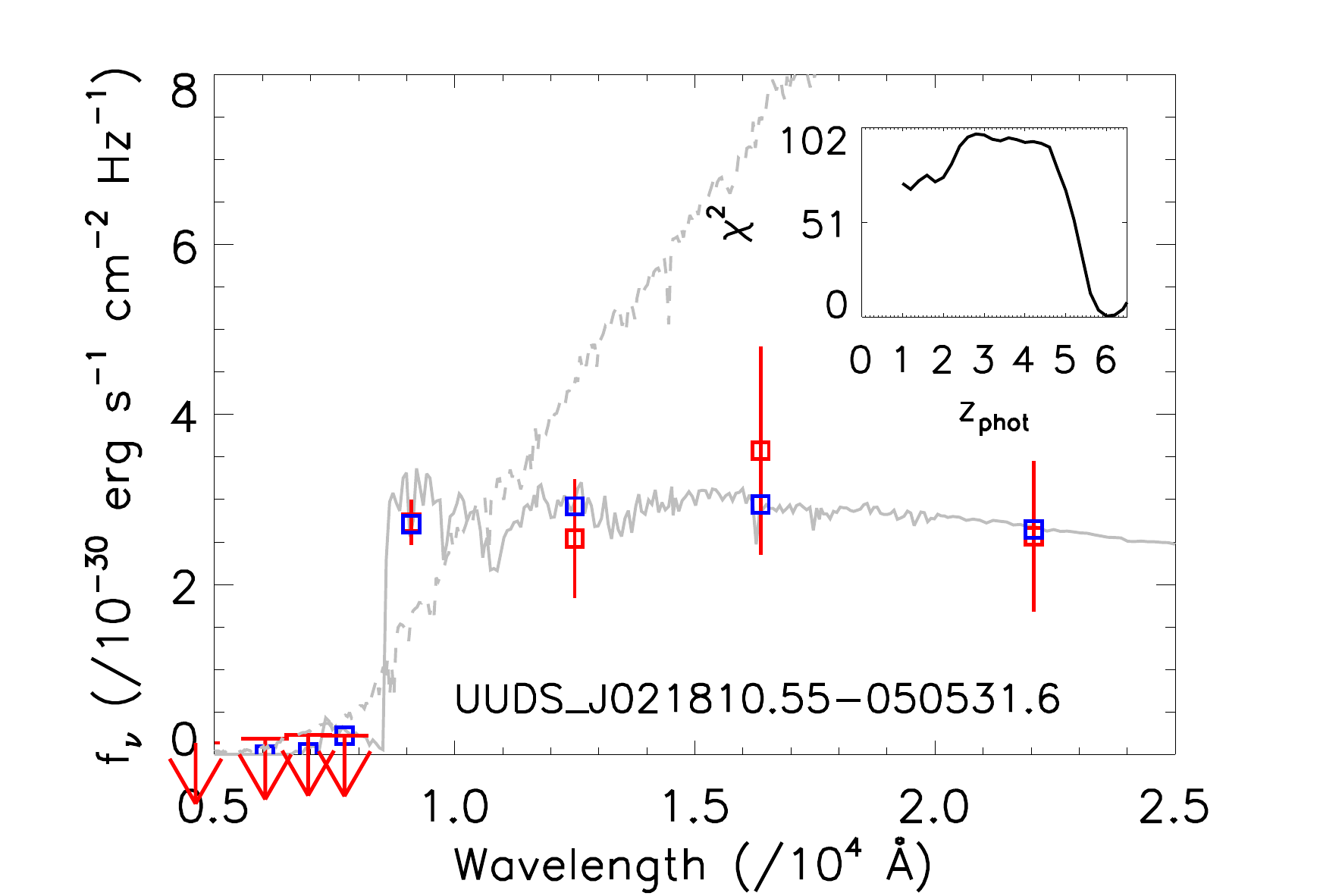}}
  \subfigure{\includegraphics[width=2.5in,trim=2.8cm 13.2cm 0.5cm 4.5cm,clip]{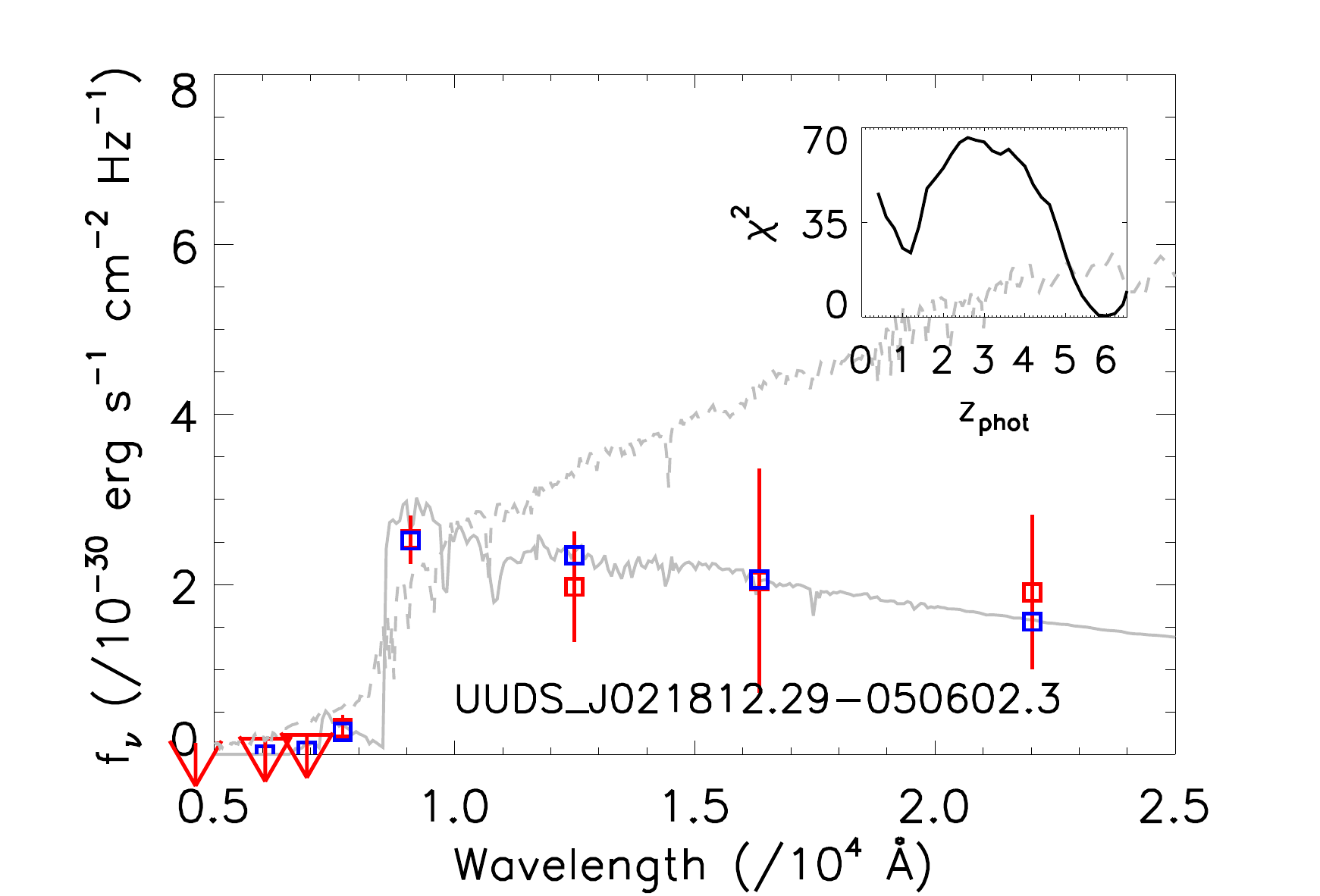}}
  \subfigure{\includegraphics[width=2.5in,trim=2.8cm 13.2cm 0.5cm 4.5cm,clip]{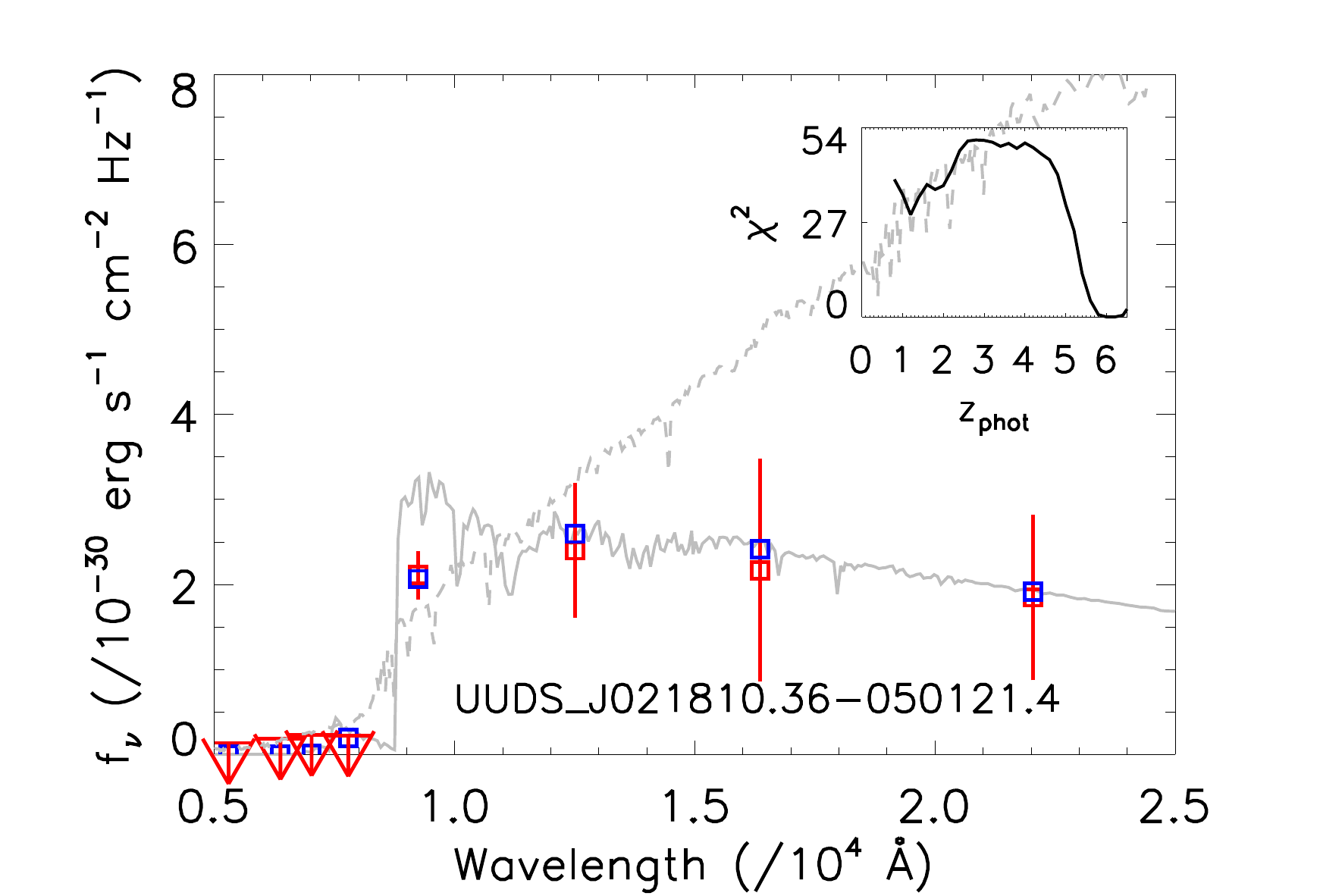}}
  \caption{SED fits for the three unconfirmed objects and UUDS\_J021816.33-051116.6, which is assigned a quality 'C' emission line redshift (see text) and is not treated as a secure Ly$\alpha$ detection.  For the three objects without an assigned spectroscopic redshift, only the primary and secondary photometric redshift solutions are shown as grey solid and dashed lines respectively.  For these objects the model fluxes are calculated from the primary redshift solution (blue points).}
  \label{fig:photoz2}
\end{figure*}

\subsubsection{Spectra of unconfirmed sources}
Four of the $z\geq6$ LBG candidates which were spectroscopically observed did not produce
spectra which allow us to assign a robust redshift. One of these four objects 
(UUDS\_J021816.33-051116.6) does display a probable Ly$\alpha$ emission line but is assigned 
a quality flag of C since the emission line sits within a region of poorly subtracted sky-line residuals. One further object does not display a prominent Ly$\alpha$ emission line, 
but does display a continuum break consistent with a redshift of $z=5.83$. However, the 
signal-to-noise in the final FORS2 spectrum is too low to be confident of the redshift. 

The spectra of the remaining two candidates show only a very faint continuum, with a signal-to-noise ratio too 
low to reliably distinguish between a Lyman break and low-redshift interloper. We note that three of the four 
unconfirmed candidates lie on the same FORS2 mask. However, although this mask is slightly poorer than average, 
in terms of redshift completeness for primary targets at $1<z<1.5$, there is no particular reason to suspect that the spectra of these three high-redshift
candidates have been unduly affected.

\subsection{Photometric redshift selection and spectroscopic completeness}
From the sample of fourteen $z\geq 6$ LBG candidates which were selected for spectroscopic follow-up, a total of ten objects have been
robustly spectroscopically confirmed as $z\geq6$ objects. As a consequence, regarding the other four spectra as contaminants, the lower limit to the 
spectroscopic completeness for our sample is 71\%. However, as described above, two of the remaining spectra display strong evidence of
being genuine high-redshift galaxies at $z\geq 5.8$ and have only failed to make our robust sample due to insufficient signal-to-noise. If we regard 
these two objects as also being confirmed $z\simeq6$ objects, then the spectroscopic completeness of our sample rises to 86\%.  Irrespective of the nature
of the two uncertain candidates, the high spectroscopic completeness provides a strong vindication of the original method of selecting high-redshift targets
on the basis of an SED-fitting photometric redshift analysis.

To illustrate this point, in Fig. 2 we show the latest photometry available for each of the spectroscopically targeted $z\geq6$ objects, 
along with the results of our updated photometric redshift analysis. For each object the observed photometry is plotted along with the best-fitting SED model returned 
when redshift is kept as a free parameter, as well as the best-fitting model using the fixed 
spectroscopic redshift and adding a Ly$\alpha$ emission line with the correct EW. 
The photometric redshifts were calculated using LePhare photometric redshift code \citep{Ilbert2006}.
The galaxy templates were made using the \cite{Bruzual2003} stellar
population synthesis models, with exponentially decreasing star
formation rates with $e$-folding times in the range 0.1~Gyr~$<\tau<$~30~Gyr and a \cite{Chabrier2003} initial mass function (IMF).
Reddening by dust is added using the \cite{Calzetti2000} extinction
law with E(B-V) values ranging from 0.0 to 0.5.   For the fitting with
fixed redshift, Ly$\alpha$ is added to the models after extinction is
applied, using the flux from the template convolved with the
narrow-band filter as the estimate of the continuum to translate the
EW into the Ly$\alpha$ flux.  

In Fig. 3 the insets showing the distribution of $\chi^2$ versus
photometric redshift clearly demonstrate, as expected, that the primary
photometric redshift solutions are all at $z \gtsim 6$.  Crucially, it
can also be seen that any competing low-redshift  solution (shown as
the dashed grey SED fits) can be ruled-out at high statistical
significance. The fundamental reason for  this is the additional
information on the spectral slope long-ward of Ly$\alpha$
provided by the UDS  near-infrared photometry. Without this additional
information any sample of bright $z\simeq 6$ LBGs candidates is
vulnerable to substantial  contamination by low-redshift interlopers,
a point we will return to in Section 4.

\section{Lyman Alpha}

\begin{figure}
  \centering
  \includegraphics[width=3in,trim=3cm 12cm 1cm 3cm,clip]{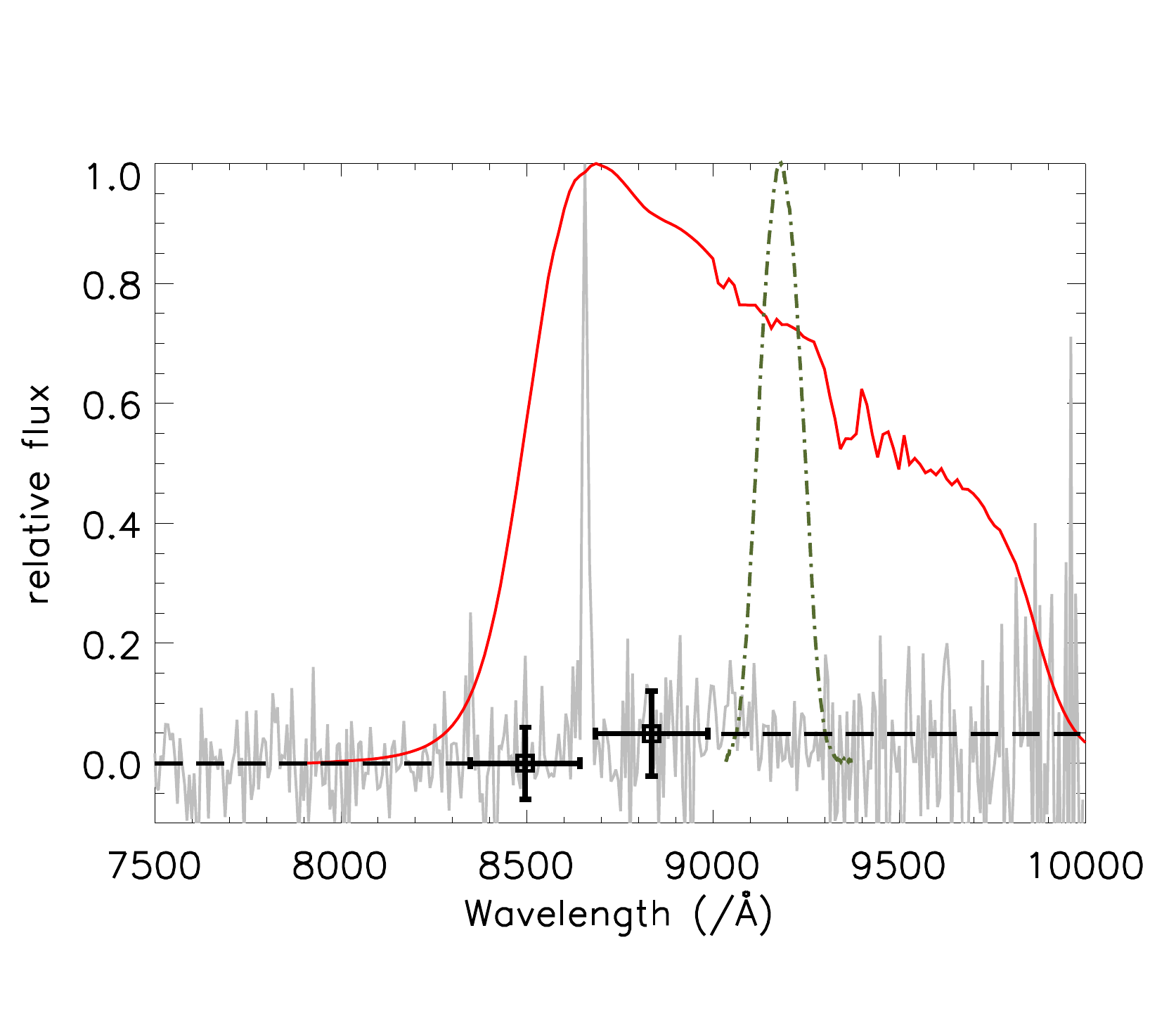}
\caption{An illustration of the method adopted to calculate the
  Ly$\alpha$ EWs. The spectrum is of object UUDSJ$\_$021735.34-051032.6 scaled to the
  height of the Ly$\alpha$ line.  Over-plotted as the red solid line is
  the throughput of the Subaru $z^{\prime}$-band filter.  The spectrum is
  convolved with this filter and scaled to match the $z^{\prime}$-band
  photometry.  To ensure that sky residuals do not dominate the
  convolution the continuum is modelled as a simple step function, as
  shown by the black long dashed line, where the continuum is derived
  from regions either side of the emission lines (shown by the points
  with associated error bars). The green dot-dashed line shows the
  Subaru narrow-band NB921 filter profile. Photometry from this filter
  is used to determine the rest-frame UV flux red-ward of Ly$\alpha$
  emission. This is used with the calibrated line flux to determine
  the Ly$\alpha$ EW.}
  \label{fig:EWprocedure}
\end{figure}  

\subsection{Ly$\alpha$ equivalent widths}

The Ly$\alpha$ equivalent widths were obtained by measuring the
observed line flux, and estimating the UV continuum at Ly$\alpha$ from
Subaru NB921 narrow-band (NB) imaging available in the UDS
\citep{SobralDavid2011}.  

The line flux was measured by subtracting any continuum from the line
profile, where the continuum is estimated from regions either side of
the line unaffected by sky line residuals, then integrating the flux
over the pixels contributing to the line.  To ensure that the
calibration of the spectra matches the photometry, they are convolved
with the Subaru $z^{\prime}$-band filter profile and the flux scaled to match the
$z^{\prime}$-band photometry (see Fig.~\ref{fig:EWprocedure}).  To ensure that
sky residuals do not dominate the derived flux from the convolved
spectrum, the continuum level is estimated just red-wards and
blue-wards of Ly$\alpha$ and is modelled by a simple step function.
Over the small wavelength range within the filter, differences in true
slope red-wards of Ly$\alpha$ produce minimal differences to the final
convolved flux.

Absolute calibration of the spectra in this way is not effective if
the signal-to-noise is so low in the continuum that it is effectively
measured as zero.  In these cases (noted in Table 1) the median
correction factor measured from all the other spectra is used and the
standard deviation of correction factors is folded into the final
error for these line fluxes.

The final error in the Ly$\alpha$ flux measurement includes pixel
error estimates from the noise array, estimated error in the
subtracted continuum measurement, as well as the estimated error in
the derived calibration.  This calibration error includes the error in
the convolved flux, assuming errors per pixel derived from the
standard deviation of pixel values for those contributing to continuum
estimates, as well as the errors in the $z^{\prime}$-band photometry.

The NB921 photometry samples the continuum red-wards of Ly$\alpha$ for
$z<6.5$ (see Fig.~\ref{fig:EWprocedure}) and to estimate the continuum
at Ly$\alpha$ a flat UV continuum in $F_{\nu}$
($F_{\lambda}\propto\lambda^{-2}$) is assumed. The NB photometry is used to estimate
the UV continuum rather than direct measurements made from the spectra
because sky subtraction hinders accurate continuum measurements in the
spectra, as well as the signal-to-noise in any detected continuum
being very low ($<2$) in many of the spectra.  The errors in the
derived EWs include the errors in NB photometry \footnote{For the two highest redshift LBGs, the Ly$\alpha$ emission line lies within the 
extreme blue-end of the NB921 filter. Although the throughput of the filter is extremely 
low at these wavelengths, it is possible that the resulting EW measurements are underestimated by $\lesssim5$\%.}. No attempt was made to correct either 
the UV continuum or Ly$\alpha$ line flux for dust extinction during the EW calculation.

Both the $z^{\prime}$-band and narrow-band fluxes were corrected to total using
measured aperture corrections from stars within the individual SXDS
pointings.  The measured equivalent widths were compared to estimates
taken directly from the spectra themselves and are found to be in good
agreement, indicating that the quoted equivalent widths are secure.

\begin{table}
 \centering
  \caption{Star-formation rates derived from Ly$\alpha$ emission-line luminosity and UV continuum luminosity (without any correction for dust extinction). The UV-derived SFR (SFR$_{UV}$) is estimated from the UV luminosity at $\lambda_{rest}=1500$\AA{ }using equation 1 and the Ly$\alpha$-derived SFR (SFR$_{Ly\alpha}$) is estimated from the luminosity of the Ly$\alpha$ emission using equation 2.}
  \begin{tabular}{@{}lcc@{}}
  \hline
  \hline
  ID & SFR$_{Ly\alpha}$ & SFR$_{UV}$\\
            & (/M$_{\odot}$ yr$^{-1}$) & (/M$_{\odot}$ yr$^{-1}$)\\
  \hline
  UUDS\_J021800.90-051137.8 & 13.8 $\pm$ 1.3 & 30.2 $\pm$ 2.6\\
  UUDS\_J021616.53-050217.7 & \phantom{0}7.4 $\pm$ 1.3 & 18.4 $\pm$ 1.7\\
  UUDS\_J021807.14-045841.5 & 12.3 $\pm$ 1.1 & 40.6 $\pm$ 3.8\\
  UUDS\_J021735.33-051032.6 & 13.4 $\pm$ 1.1 & 26.3 $\pm$ 3.6\\
  UUDS\_J021838.90-050944.0 & 16.0 $\pm$ 1.5 & 37.0 $\pm$ 3.8\\
  UUDS\_J021653.00-044623.3 & \phantom{0}4.2 $\pm$ 0.5 & 30.5 $\pm$ 2.4\\
  UUDS\_J021922.01-045536.3 & \phantom{0}7.5 $\pm$ 1.0 & 13.2 $\pm$ 2.4\\
  UUDS\_J021841.02-051247.4 & \phantom{0}8.8 $\pm$ 1.5 & 29.1 $\pm$ 4.4\\
  UUDS\_J021701.44-050309.4 & \phantom{0}5.0 $\pm$ 0.9 & 43.3 $\pm$ 4.0\\
  \hline
  \hline
\end{tabular}
\label{tab:SFR}
\end{table}

\subsection{Ly$\alpha$ and UV-derived SFRs}

\begin{figure}
  \centering
  \includegraphics[width=3in,trim=3cm 12cm 2cm 3cm,clip]{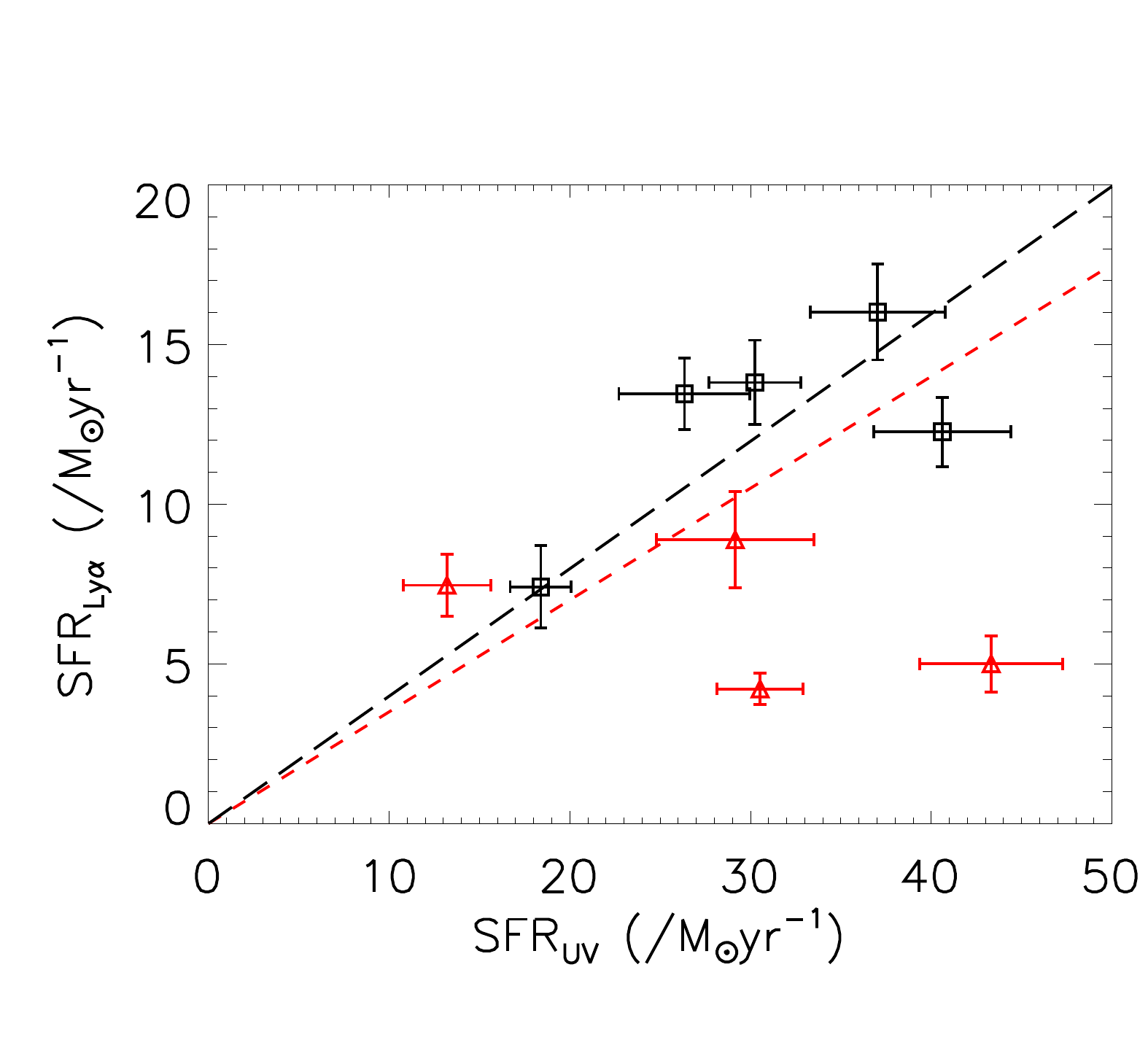}
  \caption{Ly$\alpha$-derived SFRs vs. UV-derived SFRs.  The short dashed red line shows the mean Ly$\alpha$:UV SFR ratio (0.36) and the black long dashed line shows the ratio of 0.4 predicted from Hayes et al. (2011) (for $f_{esc}^{Ly\alpha}=0.25$ at $z\sim6$).  Objects with $z>6.2$ are plotted as red triangles.}
  \label{fig:uvlyasfr}
\end{figure}

Both the UV continuum and Ly$\alpha$ emission-line luminosity are commonly used to
to derive SFR estimates. Both of these indicators are
sensitive to the presence of dust and Ly$\alpha$ is also sensitive to
neutral hydrogen within the galaxy and the surrounding IGM.
In Fig. \ref{fig:uvlyasfr} we compare the SFR estimates 
derived from the UV continuum luminosity using the \cite{Madau1998} formula (corrected to a
Chabrier IMF), with those derived from the measured Ly$\alpha$ luminosity (e.g. \citealt{Nilsson2009}), 
with no correction for dust extinction: 
\begin{equation}
L_{UV}\, \mathrm{(/erg{ }s^{-1} Hz^{-1})} = 4.8\times10^{27} \times
SFR_{UV}\, \mathrm{(/M_{\odot} yr^{-1})}
\end{equation}
\begin{equation}
L_{Ly\alpha}\, \mathrm{(/erg{ }s^{-1})} = 9.7\times10^{41}\times
SFR_{Ly\alpha}\, \mathrm{(/M_{\odot} yr^{-1})}
\end{equation}
\noindent
The constant in equation 2 is based on the continuum luminosity at
$\lambda_{rest}=1500$\AA{ }and has been converted to a Chabrier IMF using a ratio of 1.65, the asymptotic ratio between the number of ionizing photons produced by a
constant SFR with a Salpeter compared to a Chabrier IMF predicted by \cite{Bruzual2003}.

Although there is a large scatter between the two SFR measurements, it
can immediately be seen from Fig. 5 that the SFR estimate provided by
the Ly$\alpha$ luminosity is systematically lower than that provided
by the  UV continuum luminosity, with a mean value of SFR$_{Ly\alpha}$/SFR$_{UV}=0.36\pm0.05$, and a median of 0.4.

We note that this is very close to the value of 0.4 which would be predicted by the recent study of the redshift evolution of the
 Ly$\alpha$ escape fraction by \cite{Hayes2011}. At low redshift \cite{Hayes2011} calculate $f_{esc}^{Ly\alpha}$ by comparing the observed
 Ly$\alpha$ luminosity function with a prediction based on the extinction corrected H$\alpha$ luminosity
function (assuming case B recombination). The lack of observed H$\alpha$ luminosity functions above $z\sim2.3$, means that at higher redshifts, the intrinsic
Ly$\alpha$ luminosity function is predicted from the UV, with extinction estimates being made from SED fitting.  
This calculation is essentially taking the ratio of the Ly$\alpha$ and UV star formation rate densities, $\dot{\rho}_{\bigstar}$ 
(equations 3-5), where the star-formation rate densities are derived from integrating over the
Ly$\alpha$ and UV luminosity functions respectively, i.e.  

\begin{equation}
f_{esc}^{Ly\alpha}=C_1\left(\frac{\dot{\rho}_{\bigstar\, Ly\alpha}^{Obs}}{\dot{\rho}_{\bigstar\, UV}^{Obs}}\right)^{1+C_2}
\end{equation}
\begin{equation}
C_1={C_{Ly\alpha}}^{-C_2}
\end{equation}
\begin{equation}
C_2=\frac{k_{UV}}{k_{Ly\alpha}-k_{UV}}
\end{equation}
\noindent 
Where $k_{UV}=10.3$ (Calzetti et al. 2000) and $k_{Ly\alpha}=13.8$ and $C_{Ly\alpha}=0.445$ are derived from a fit to the observed anti-correlation between E$_{\mathrm{B-V}}$ and $f_{esc}$. Based on this formalism, \cite{Hayes2011} find that the escape fraction varies as 
$f_{esc}^{Ly\alpha}\propto(1+z)^{\xi}$ with $\xi=2.57^{+0.19}_{-0.12}$, giving $f_{esc}^{Ly\alpha}\simeq0.25$ at $z\simeq6$, hence providing an estimate of SFR$_{Ly\alpha}$/SFR$_{UV}\sim0.4$, which agrees with the mean value derived here, within the errors. Consequently, based on the results of \cite{Hayes2011}, we conclude that our sample of $L\geq 2L^{\star}$ LBGs at $z\simeq6$ are consistent with a Ly$\alpha$ escape fraction of $f_{esc}^{Ly\alpha}\simeq 25\%$.


Finally, we note that in Fig. 5, while most of the sample agree well with the prediction of \cite{Hayes2011}, the two largest
outliers are drawn from the high-redshift end of the sample at $z>6.2$ and suggest a lower value of $f_{esc}^{Ly\alpha}$.

\section{The Lyman-alpha emitter fraction}
\begin{figure}
  \centering
\includegraphics[width=6.5in,angle=0,trim=3cm 3cm 1cm 6cm, clip=true]{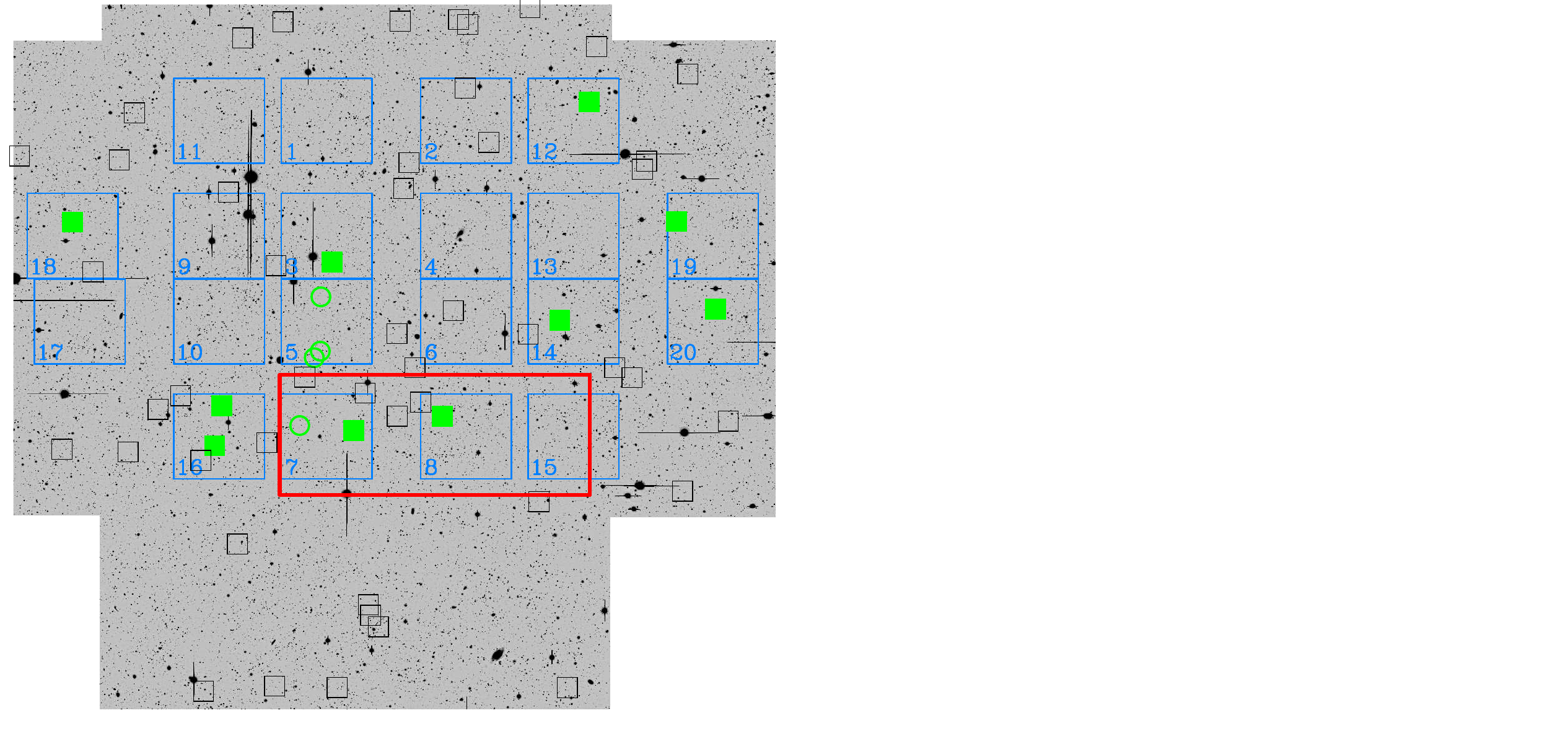}
    \caption{A greyscale representation of the region with overlapping
    Subaru (optical) and UKIRT (near-infrared) data in the UDS.  Over-plotted
         in blue are the locations of the 20 FORS2 masks from the UDSz
     programme and the area imaged by HST/WFC3 as part of the CANDELS survey (Grogin et al. 2011) 
is shown in red.  The small black squares show all the
    objects in the field that are plausible $z\geq6$ candidates following
        the criteria described in the text. The green squares show the
  spectroscopically confirmed $z\geq6$ LAEs, while the green circles show
             the objects for which the FORS2 spectra do not provide an
                                     unambiguous redshift.}
  \label{fig:sample}
\end{figure}
 In this section we use our sample of spectroscopically confirmed LBGs
to provide entirely new information on the Ly$\alpha$ emitter fraction
 of $L\geq 2L^{\star}$ LBGs at $z\geq 6$. As previously described, the
    original sample of spectroscopic candidates was selected using the
       redshift probability density function derived using a photometric
        redshift analysis (McLure et al. 2009). However, although this
 strategy makes optimal use of the available data, in order to perform
    a comparison between our results and those in the literature it is
          necessary to re-engineer our candidate selection in terms of
      traditional colour-cut criteria. Consequently, all of the parent
sample considered for spectroscopic follow-up also satisfy the following criteria:
\begin{equation}
z' < 26
\end{equation}
\begin{equation}
i'-z'\geq 2.0
\end{equation}
\begin{equation}
z'-J \leq 0.8
\end{equation}
\noindent
where all magnitudes are measured within a 2\asec{ }diameter
aperture and we also require each object to be a non-detection in the $BVR$
bands at the $2\sigma$ level. The choice of a stringent $i'-z'\geq 2$
colour cut is motivated by the fact that any galaxy at $z\geq 6$ is
predicted to  display a drop of at least two magnitudes between the
Subaru $i'$ and $z^{\prime}$ filters assuming the Madau (1995) prescription
for IGM absorption.  The $z'-J\leq 0.8$ colour-cut is motivated by the
desire to effectively exclude low-redshift interlopers at $z\simeq
1.5$, while maintaining  sensitivity to genuine $z\geq 6$ galaxies
with moderate reddening.

Within the overlap region between the optical and near-infrared imaging in the UDS the surface density
of objects obeying the above selection criteria is $0.023\pm0.003$ per square arcmin 
Therefore, within the total area covered by the FORS2 pointings (20 pointings, each covering an area of 6.8$^{\prime}\times6.8^{\prime}$), 
we would expect $\sim15-20$ candidates. From Fig.~\ref{fig:sample}, which shows the region of optical/near-infrared overlap in the UDS
with the positions of the FORS2 masks and the candidates satisfying the above criteria over-plotted, it can be seen that there are 19 candidates in this
area, with 14 of those having been targeted for spectroscopy. Consequently, we conclude that the sample of candidates 
targeted for spectroscopy appears to be consistent with being drawn randomly from the parent population of candidates satisfying the 
colour-cut criteria described above (although see section 4.2.4 for further discussion).

From our sample of fourteen objects we find seven which display strong Ly$\alpha$ emission (EW$\geq$25\AA), which suggests a minimum
Ly$\alpha$ emitter fraction of 50\%. However, if we exclude the AGN from our sample this fraction increases to 54\%. 
Either way, these are very high Ly$\alpha$ fractions compared to previous literature results for luminous LBGs and 
clearly require some discussion.

\subsection{Comparison with the literature}

\begin{figure}
  \centering
  \includegraphics[width=3in,trim=3cm 13cm 1cm 3cm,clip]{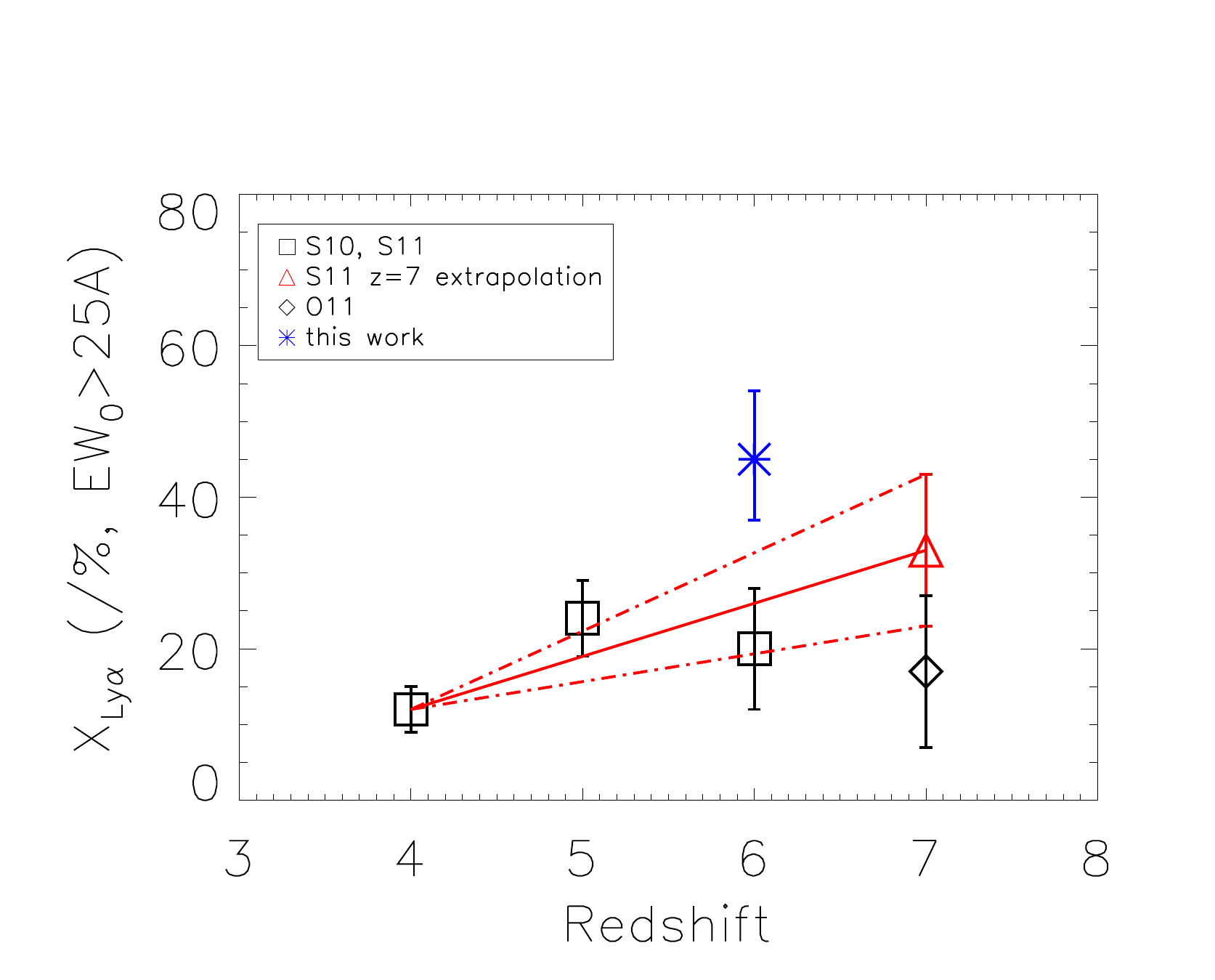}
  \caption{The evolution with redshift of the fraction of luminous LBGs ($-21.75<$ M$_{UV}<-20.25$) which display Ly$\alpha$ line emission with rest-frame EW$\geq 25$\AA . The black squares show the measured fractions from Stark et al. (2010, 2011) and the black diamond is the measured fraction at $z\sim7$ estimated by Ono et al. (2011) from the combination of a number of spectroscopic samples.  The red triangle shows the expected value at $z\sim7$ extrapolated from the trends at lower redshift (Stark et al. 2011), with the fitted linear relationship and uncertainties plotted as the red solid and dashed lines, respectively.  The blue star shows the results of this work based on the
sample selected from the UV-continuum flux and with upper and lower limits as described in the text.}
  \label{fig:frac}
\end{figure}

We now look to other studies of spectroscopically targeted LBGs at
$z\sim6$ to see whether this Ly$\alpha$ fraction is comparable.  \cite{Stark2011} present a sample of
$i'$-band dropouts selected with a
$i_{775}-z_{850}>1.3$\footnote{$i_{775}$ and $z_{850}$ are the F775W and F850LP HST/ACS
  filters respectively.} colour cut, which is a considerably bluer cut than that adopted here.  
This sample, when compared to results at lower redshift \citep{Stark2010}, provides strong evidence that the fraction of objects
showing Ly$\alpha$ in emission increases from $z=4$ to $z=6$ for galaxies in the absolute magnitude range $-20.25<M_{UV}<-18.75$. However, in contrast, 
little evolution in the Ly$\alpha$ fraction is seen for brighter objects with UV luminosities more comparable to our sample ($-21.75<M_{UV}<-20.25$).
In this luminosity range, Stark et al. (2011) find that the fraction of objects
showing Ly$\alpha$ in emission is $20\% \pm 8.1\%$ and $7.4\% \pm 5.0\%$ for objects with EWs
greater than 25\AA{ }and 55\AA{ }respectively.

Considering the small size of our sample, at the higher EW$\geq55$\AA\, threshold
our results are in reasonably good agreement with \cite{Stark2011}, given that one of our
objects has EW $\simeq55$\AA{ }(i.e. 1/9).
However, at the lower EW$\geq25$\AA\, threshold our sample shows a substantially larger fraction of LBGs with
Ly$\alpha$ emission ($\simeq54\%$). 

\subsection{Potential biases}
Understanding or reconciling the difference in the measurements made
by these two studies is difficult due to the different selection mechanisms used 
and the small numbers of bright objects available for study in fields smaller than the UDS.
In fact, it is not clear that a direct comparison between our results and those of \cite{Stark2011} is actually meaningful, given 
that the vast majority of our sample are confined to the bright half of the luminous absolute magnitude 
bin ($-21.75<M_{UV}<-20.25$) defined by \cite{Stark2011}. However, given that the trend reported by Stark et al. (2011), 
is for a {\it decreasing} fraction of strong Ly$\alpha$ emitters at $z\simeq6$ with increasing UV luminosity, our results indicating a 
high fraction of Ly$\alpha$ emitters amongst luminous $z\geq 6$ LBGs clearly requires some explanation. In this section we 
investigate several effects which could have potentially biased our determination of the Ly$\alpha$ emitter fraction.

\subsubsection{Reddening}
First, we consider whether our colour selection criteria are selecting
against galaxies with internal reddening by dust that may produce
lower Ly$\alpha$ EWs. The $i'-z'\geq 2.0$ colour cut does not select against reddened galaxies as this colour 
is dependent on the fraction of flux attenuated in clumpy HI regions in the intervening IGM,
which is independent of the galaxy properties. Reddening tends to increase this colour since there is more flux at 
redder wavelengths still contributing to the $z'-$band flux.  It is possible that the true IGM absorption for individual 
galaxies may scatter the colours of genuine $z\geq 6$ galaxies below the colour cut, but this scatter should be independent
of the intrinsic galaxy properties.  The $z'-J$ colour cut may well
exclude a few redder $z\geq 6$ galaxies from our sample. This colour cut is designed to exclude low-redshift galaxies and 
cool galactic stars which are able to satisfy the $i'-z'$ colour cut. However, because we are insisting on a strict $i'-z'\geq 2.0$ 
colour, only five additional objects are excluded on the basis of the $z'-J\leq 0.8$ criterion over the entire UDS field, so we do not expect our 
sample to be greatly biased towards dust-free galaxies.

However, it is important to note that if a bluer $i'-z'$ colour cut is adopted, the potential for contamination increases due
to a population of massive red galaxies at $z\simeq 1.5$ with a surface density which increases 
significantly at $z'\leq 26$. As a result, any samples selected without near-infrared data and a bluer 
$i'-z'$ colour cut will be significantly contaminated by low-redshift interlopers at $z'\leq 26$. Of course, until the present study, 
selecting significant samples of $z\simeq 6$ objects with $z'\leq 26$ has not been possible due to insufficient area.

\subsubsection{Contamination from low-redshift interlopers}
To further explore how low-redshift interlopers could contaminate samples of bright $z\simeq 6$ LBGs, we have investigated 
different selection criteria using the deep HST data available in GOODS-S.
All of our objects were found to satisfy the HST colour selection of
$i_{775}-z_{850}>1.7$.  This lower colour cut is due to the difference
in filter profiles between the HST $z_{850}-$band compared to the Subaru $z^{\prime}$
band, as well as slightly increased overlap between the HST $i_{775}$ and $z_{850}$ band filters.  
With the new HST/WFC3 near-infrared data available as part of the CANDELS survey
\citep{GroginNormanA.2011,KoekemoerAntonM.2011} we can investigate how
our $z'-J$ cut may affect these samples. We find that only two objects in the 
field satisfy $i_{775}-z_{850}>1.7$ with $z_{850}<26$, {\it both} of which are excluded by a $z_{850}-J_{125}<0.75$ criterion 
(consistent with our $z'-J<0.8$ colour cut). However, in a sample selected using $i_{775}-z_{850}>1.3$; $z_{850}<27$, we find that
$>90$\% of objects survive the $z_{850}-J_{125}<0.75$ colour cut. It is therefore clear that while a colour-cut of $i_{775}-z_{850}>1.3$ should be sufficient to select a clean sample of $z\simeq 6$ LBGs at $z_{850}\simeq 27$, at magnitudes brighter than $z'\leq 26$ suitably deep near-infrared photometry is essential to avoid significant contamination from low-redshift interlopers. In fact, this point is very 
well illustrated by the SED template fits to our spectroscopically confirmed LBGs shown in Fig. 2.

\subsubsection{EW measurements and cosmic variance}
It is worth noting at this point that two of our objects with
strong Ly$\alpha$ emission have EWs that are quite close to the
EW$\geq25$\AA{ }cut.  Although our method of EW calculation is quite
robust, not all fields have narrow band imaging just red-wards of 
Ly$\alpha$ to supply a continuum estimate, and it is difficult to get
high signal to noise in the continuum from the actual spectra at
these redshifts.  It is certainly feasible that with a different
continuum estimator, or a lower limit derived from the noise in the
spectrum, that these EWs could be measured to be lower, pushing down
the measured Ly$\alpha$ fraction.  

Moreover, it is also important to consider the problem of cosmic variance. 
Even selecting over an area of 0.25 sq. degrees, our sample is clearly limited by
small number statistics. As illustrated by Fig. 6, previous samples of luminous $z\sim 6$ LBGs 
selected over the GOODS fields (the CANDELS imaging in the UDS covers the area of one GOODS field) will
include very few, if any, objects as luminous as those in our sample. Again, this simply highlights 
the point that our sample is exploring a new area of parameter space.

\subsubsection{Photometric redshift selection}
One potential issue which needs to be addressed is the fact that our original object selection was performed using the
Subaru $z^{\prime}$-band imaging in the UDS which, for objects at $z\geq6$, is affected by both IGM absorption and a varying Ly$\alpha$ emission line contribution. As a result, we can probably regard our estimate of 54\% for the fraction of
luminous $z\simeq6$ LBGs which display Ly$\alpha$ emission with EW$\geq25$\AA{ }as an upper limit. 
The fundamental reason for this is
that, due to our original photometric redshift selection our
spectroscopic sample is not an entirely random sampling of the parent
population satisfying the colour-cut criteria listed previously.  In
fact, because we targeted objects with the highest probability of
being at $z\geq6$, the location of the Ly$\alpha$ emission line within 
the selection filter introduces a bias towards preferentially
targeting objects with strong Ly$\alpha$ emission. In short, the
presence of strong Ly$\alpha$ emission leads to an exaggerated Lyman
break and an apparently bluer UV spectral slope, both of which
can conspire to produce a more robust photometric redshift solution at
$z\geq6$ than would otherwise be the case.

\subsection{An unbiased estimate of the Lyman-alpha fraction using UV continuum selection}
Given the complications introduced by selecting our original $z\geq 6$
sample in the Subaru $z'-$band, we can take advantage of the
availability of the NB921 imaging data to investigate the Ly$\alpha$
fraction in a UV-continuum selected sub-sample. In general, for
$z<6.4$ where the narrow-band imaging is clear of the Lyman-break, we
expect the narrow-band fluxes to be the same or brighter than the
$z'-$band fluxes for  any object with no Ly-$\alpha$ emission (this is
reversed if there is strong Ly$\alpha$ emission contributing to the
$z^{\prime}$-band). A comparison of the $z'-$band and NB921 photometry
for our sample suggests that our original  selection limit of $z'\leq
26$ provides a complete sample of objects with $z_{921}\leq 25.7$
(all objects satisfying equations 7 \& 8 with $z_{921}\leq 25.7$ also
obey $z'\leq 26$). This  narrow-band magnitude cut corresponds to an
absolute UV magnitude of M$_{UV} = -21.0$ at $z=6$ and M$_{UV}=
-21.1$ at $z=6.5$ (assuming a flat UV slope in $f_{\nu}$).  At this
limit we find 45\% of objects (5/11) with Ly$\alpha$ EW
$\geq25$\AA, confirming a high fraction of Ly$\alpha$ emitters at
M$_{UV}\leq-21.1$.

\subsection{A lower limit to the Lyman-alpha fraction}
Using our spectroscopic sample it is also possible to estimate a hard
lower limit to the Ly$\alpha$ fraction amongst luminous $z\geq6$
LBGs, if  we consider the unlikely scenario in which our EW$\geq
25$\AA\, Ly$\alpha$ detections are the {\it only}  strong Ly$\alpha$
emitters out of the 19 objects which could have been spectroscopically
targeted. Even in this extreme scenario the fraction of galaxies
showing Ly$\alpha$ with EW$\geq$25\AA{ }is $37$\%$\pm$14\%
(7/19), where the quoted error is the poisson uncertainty.

However, in reality, this situation is better described by a binomial 
distribution which predicts that the probability of finding 7 or more galaxies with EW $\geq$ 25\AA\, 
from a sample of 14 is only $p=0.01$, if the chance of ``success'' in each trial (i.e. finding EW $\geq 25$\AA) 
is $p=0.20$ (as is the case for the most luminous $z\sim6$ objects in Stark et al. 2011). However, given the potential for bias, 
perhaps a more reasonable calculation is to determine the probability of finding 7 or more galaxies with EW $\geq 25$\AA\, from 
all 19 {\it potential} spectroscopic targets. This returns a probability of $p=0.07$, meaning that we are only able to 
exclude the hypothesis that the true fraction of EW$ \geq$ 25\AA\, Ly$\alpha$ emitters within our sample is 20\% at the $\simeq 93\%$ confidence level.

\subsection{Dust free luminous LBGs}
Our Ly$\alpha$ fraction results are plotted in Fig. 7 along with the results of Stark et al. (2010, 2011) and Ono et
al. (2011). It is clear from this plot that our best estimate of
$\simeq 45\%$ for the Ly$\alpha$ fraction amongst luminous LBGs at
$z\simeq 6$ is significantly higher than found by Stark et al. (2011)
at the same redshift and by Ono et al. (2011) at $z\simeq 7$ (although
we note that, before combining their results with other
studies taken from the literature, they find a fraction of $\simeq
33\%$). As far as we can tell, this result is not significantly biased
by our sample selection and, as can be seen from Fig. 7, is in good
agreement with a simple extrapolation of the Ly$\alpha$ fraction
derived by Stark et al. (2010) for luminous LBGs at $z=4$ and $z=5$.

One possibility is that by sampling the bright end of the $-21.75<M_{UV}<-20.25$
bin, we are starting to see the effects of targeting the extreme tail of the population of the less-dusty, highly star 
forming systems.  As we move to brighter and brighter objects, the extremely steep slope of the luminosity function 
would seem to require that samples eventually become dominated by objects with low dust reddening and, presumably, a 
correspondingly high escape fraction of Ly$\alpha$ photons.  Although this scenario would run counter to the observation that, at a given redshift, brighter UV-selected galaxies are generally redder than their low-luminosity counterparts, we note that the median UV slope of our sample is $\langle\beta\rangle=-2.51$.  This is bluer than the median value of $\beta=-2.06^{+0.12}_{-0.09}$ for $z\sim6$, $L>0.75L^{*}$ derived by \cite{FinkelsteinStevenL.2011} and the variance-weighted mean value derived at $z\simeq 6.5$ by McLure et al. (2011) of $\langle\beta\rangle=-2.05\pm0.09$, although in the absence of deeper near-IR photometry it is difficult to draw any firm conclusions regarding the UV slopes of our sample.

\subsection{A UV continuum selected sample}
As discussed previously, selecting $z\geq 6$ galaxy samples based on $z^{\prime}-$band photometry involves
the added complications of varying amounts of IGM absorption and Ly$\alpha$ contamination.
In this study, we were able to correct any possible biases by using narrow-band imaging data red-wards of Ly$\alpha$ to
define a sample complete to a given depth in the UV continuum.  However, to achieve this at higher redshifts it is clearly necessary to perform
the primary sample selection in the near-infrared. Within this context, the 
new Cosmic Assembly Near-infrared Deep Extragalactic Legacy Survey (CANDELS; Koekemoer et al. 2011; Grogin et al. 2011) 
will supply deep, near-infrared, imaging (5$\sigma$ depth of $H_{160}\simeq27$) over a total area of $\sim700$ square arcmin.  The combination of depth and area provided by CANDELS
will allow the Ly$\alpha$ fraction amongst large, unbiased, samples of $z>7$ LBGs to be studied in the near future.

\section{Conclusions}
Targeting a sample of $z\geq6$ galaxy candidates originally selected using a photometric redshift analysis, we
obtained a very high redshift completeness, with 11/14 objects providing robust spectroscopic redshifts from 
the detection of a Ly$\alpha$ emission line. Comparing the star-formation rate estimates based on 
UV continuum and Ly$\alpha$ luminosity we find that our sample is consistent with a Ly$\alpha$ escape 
fraction of $f_{esc}^{Ly\alpha}\simeq 25\%$, in agreement with the recent study of \cite{Hayes2011}.

Based on our sample of $L\geq 2L^{\star}$ LBGs at $z\simeq6$ we derive estimates for the maximum (54\%$\pm$20\%) and minimum (37\%$\pm$14\%) 
fraction of Ly$\alpha$ emitters with EW$\geq25$\AA.
Testing whether these fractions are biased by our $z^{\prime}$-band selection criteria, we use the available NB photometry to 
calculate the fraction of EW$\geq25$\AA\, Ly$\alpha$ emitters amongst a complete, UV continuum selected, sub-sample. This 
calculation returns a fraction of $45\%\pm$15\%, consistent with our previous estimates.

Our estimate of the EW$\geq25$\AA\, Ly$\alpha$ emitter fraction in the magnitude range 
$-21.75<M_{UV}<-21.25$ is a factor of $\simeq 2$ larger than previous estimates (i.e. $20\pm8\%$; Stark et al. 2011). Indeed, we
calculate that our sample is inconsistent with a Ly$\alpha$ emitter fraction of 20\% at the $\simeq 93\%$ confidence level. However, we also note that our sample explores a higher redshift and luminosity range than previous studies. 

In summary, our results suggest that, as the epoch of reionization is approached, it is plausible that 
the Ly$\alpha$ emitter fraction amongst luminous ($L\geq 2 L^{\star}$) LBGs shows a similarly sharp increase to 
that observed in their lower-luminosity ($L\leq L^{\star}$) counterparts.

\section*{Acknowledgments}
The authors would like to acknowledge the anonymous referee for a report which
helped to improve the original manuscript. ECL would like to acknowledge financial support from the UK Science
and Technology Facilities Council (STFC) and the Leverhulme Trust.
RJM would like to acknowledge the funding of the Royal Society via the
award of a University Research Fellowship and the Leverhulme Trust via
the award of a Philip Leverhulme research prize. 
JSD acknowledges the support of the Royal Society via a Wolfson Research Merit award, and
also the support of the European Research Council via the award of an
Advanced Grant. MC acknowledges the award of an STFC Advanced Fellowship. DPS acknowledges support from NASA
through Hubble Fellowship grant \#HST-HF-51299.01 awarded by the
Space Telescope Science Institute, which is operated by the Association of Universities for Research in
Astronomy, Inc., for NASA under contract NAS5-26555. HJP, RC, and EB acknowledge
the award of an STFC PhD studentships.  WGH acknowledges the award of an STFC PDRA.  
The authors would like to thank the HiZELS team for supplying the Subaru NB921 imaging data and 
the staff at UKIRT for operating the telescope with such dedication under difficult circumstances.

\bibliographystyle{mn2e}
\bibliography{Lyman-alpha_UDS_revised2.bbl}
\label{lastpage}

\end{document}